%% file: main.tex
\documentclass[phd,cdt,leftchapter,parskip,twoside,table
]{infthesis}

\usepackage{natbib}
\usepackage{pdfpages}
\usepackage{stylefile}

\usepackage{subcaption}

\usepackage{xcolor}
\usepackage{booktabs}
\usepackage{lscape}

\usepackage{wrapfig}

\usepackage{multicol}

\usepackage{prftree}

\usepackage{bookmark}

\usepackage{microtype}

\usepackage{tocbasic}

\DeclareTOCStyleEntry[%
,beforeskip=0pt plus .2pt%
,dynnumwidth=true%
,numsep=5pt%
,indent=6.97em%
,entryformat=\footnotesize\textit%
,pagenumberformat=\footnotesize\textit%
,onstartsamelevel=\vspace{-2pt}%
]{tocline}{subsubsection}

\makeatletter
\newcommand\notsotiny{\@setfontsize\notsotiny\@ixpt\@ixpt}
\makeatother

\usepackage{soul}

\DeclarePairedDelimiter\norm{\lVert}{\rVert}%

\newcommand\Chapter[2]{
	\chapter[#1]{#1\\[0.5ex]\Large\itshape#2}
}

\newcommand\fooline{\textcolor{seabornlightblue}{\rule{\linewidth}{1mm}}}

\newcommand{\StanPP}{Stan\nolinebreak\hspace{.02em}\raisebox{.4ex}{\tiny\bf +}\nolinebreak\hspace{-.10em}\raisebox{.4ex}{\tiny\bf +}}
\def\StanPP{{Stan\nolinebreak[4]\hspace{.02em}\raisebox{.4ex}{\tiny\bf ++}}}

\DeclareCaptionFormat{listing}{\colorbox{seabornlightblue}{\parbox{\dimexpr\linewidth-2\fboxsep+1pt\relax}{#1#2#3}}}
\titleformat*{\subparagraph}{\itshape}

\title{Program Analysis of \\Probabilistic Programs}
\author{Maria I. Gorinova}

\abstract{%
	
	Probabilistic programming is a growing area that strives to make statistical analysis more accessible, by separating probabilistic modelling from probabilistic inference. In practice this decoupling is difficult. The performance of inference methods is sensitive to both the underlying model and the observed data. Different inference techniques are applicable to different classes of models, have different advantages and shortcomings, and require different optimisation and diagnostics techniques to ensure robustness and reliability.  
	
	No single inference algorithm can be used as a probabilistic programming back-end that is simultaneously reliable, efficient, black-box, and general. Probabilistic programming languages often choose a single algorithm to apply to a given problem, thus inheriting its limitations. While substantial work has been done both to formalise probabilistic programming and to improve efficiency of inference, there has been little work that makes use of the available program structure, by formally analysing it, to better utilise the underlying inference algorithm. My thesis is that it is possible to improve probabilistic programming using program analysis, and I present three novel techniques (both static and dynamic), which analyse a probabilistic program and adapt it to make inference more efficient, sometimes in a way that would have been tedious or impossible to do by hand. 

	Part I of the thesis focuses on static analysis and gives the first formal treatment of the popular probabilistic programming language Stan. While efficient, Stan constrains the space of programs expressible in the language. Programs must be written according to Stan's block syntax, which reduces compositionality. In addition, Stan does not support the explicit use of discrete parameters. Part I introduces the probabilistic programming language SlicStan: a compositional, self-optimising version of Stan, which supports both discrete and continuous parameters. SlicStan uses information flow analysis and type inference to capture conditional independence relationships in the program and transform it for inference in Stan.	The result can be seen as a hybrid inference algorithm, where different parameters are inferred according to different inference algorithms for efficiency. 
	
	Part II shows an example of dynamic analysis. The performance of inference algorithms can be dramatically affected by the parameterisation used to express a model. It is difficult to know in advance what parameterisation is suitable, as it depends on the properties of the observed data. This part demonstrates that reparameterisation can be automated by combining effect handlers in the probabilistic programming language Edward2, with variational inference preprocessing that searches over a space of possible parameterisations. 
}

\begin{document}

\begin{preliminary}

\maketitle

\begin{laysummary}

Anna and Ben are about to play a game. Anna flips two coins without showing Ben the result. Ben guesses whether each of the coins is heads or tails. If he guesses both correctly, he wins. Otherwise, Anna wins. But they both agree this game is a little unfair: after all, there is only $25\%$ chance that Ben guesses both coins correctly! 
Anna flips the coins. Attempting to make the game more fair, she gives Ben some additional information before he makes his guess: one of the coins is tails. What are Ben's chances of guessing the other coin correctly? Is the game fair now? 

Human intuition is notoriously misleading when it comes to such problems. And yet, we are faced with problems of this kind often and across disciplines. From a doctor deciding on a treatment based on their patient's symptoms and test results, to studying the universe based only on its projection on the night sky, we need ways to make inferences about unobservable phenomena given only some incomplete related data. 

Thankfully, we can express such questions about the uncertainty of phenomena using probabilistic programming languages. Such languages aim to facilitate reasoning about probabilities and making inferences. However, statistical inference is not an easy task, and so probabilistic programming has been only partially successful in practice. 

This dissertation shows that we can exploit the rich structure provided by a probabilistic program to improve inference and to make probabilistic programming more accessible. Program analysis techniques have long been used to improve conventional programming by automatically detecting errors, optimising the program so it makes better use of available resources, making it faster or more robust, or proving properties such as correctness and safety. My thesis is that program analysis techniques can also be adapted to probabilistic programming. I show three ways in which program analysis techniques improve automatic inference: one automates an optimisation task that was previously the responsibility of the user; one analyses dependencies between variables in the program to generate an efficient hybrid inference algorithm; and one performs a program re-write that improves inference robustness and efficiency, and which would have been impossible to do by hand. 

As for Ben, he is in luck. Anna accidentally swung the odds in his favour, and he has more than $66\%$ chance of winning if he guesses the remaining coin is heads.
\end{laysummary}

\begin{acknowledgements}	
\small

Many people helped along the way of completing this work, by teaching me, mentoring me, inspiring me, collaborating with me, and supporting me both academically and socially.

\textbf{Andy Gordon} has been an invaluable influence during the past years. He supported this work every step of the way and was a bottomless source of new ideas. I am grateful for everything I learned from Andy, for his patience when I was stubborn, and for letting me do things my own way.
Thank you, to \textbf{Charles Sutton}, for always being excited about this work. Charles has been the integral machine learning wisdom source during the past years and helped assess and filter a lot of ideas. I am grateful for his continuing support, both technical and personal.  
Thank you, Andy and Charles, for supervising me during the PhD, for your guidance, for the insightful discussions, and for shaping me as a researcher.  

I'm thankful to \textbf{Bob Carpenter}, my co-author \textbf{Matthijs V\'ak\'ar}, and the Stan team for teaching me a lot about Stan, for being incredibly kind, and for welcoming me to the field. Thank you to \textbf{Michael Betancourt} for his excellent lectures and tutorials on statistics, and for being a healthy source of scepticism during this journey.
Thank you to my co-authors \textbf{Dave Moore} and \textbf{Matt Hoffman}, and to the Bayesflow team, for fruitful discussions and for the excitement about probabilistic programming. I learned a lot about approximate inference from them, as well as about the design of deep probabilistic programming languages.
\textbf{Iain Murray} helped shape the course of the PhD project during annual reviews and offered regular feedback and advice. 
I loved having chats with \textbf{Ohad Kammar}, who offered both technical and professional advice, and with \textbf{Daniel Hillerström}, who was an incredible source of knowledge about effect handlers. Ohad and Daniel both read parts of this thesis and offered helpful feedback.
Thank you to \textbf{Alex Lew} for the discussions about probabilistic programming and Gen, to \textbf{Kai Xu} for helping with questions about Turing, and to \textbf{Feras Saad} for his suggestions around a better discussion of symbolic languages, which all improved this dissertation.  

A special thank you to \textbf{George Papamakarios}, who I learned so much from. I'm grateful for the time he spent offering expertise; we've had many discussions, some of which inspired parts of this work. But most of all, I'm grateful to him for his continuous support and encouragement, for being an example of integrity and fairness, and for always inspiring assertiveness.  

Thank you to my family and friends, for always believing in me, for being there in good and bad, keeping my spirit high through blizzards, winter storms, toilet roll shortages, and a global pandemic. I could not have made it without \textbf{Janie}'s sarcastic memes, \textbf{Christine}'s supermassive black hole cakes, \textbf{Alex}'s sass and cynicism, \textbf{Deena}'s gorgeous cocktails, \textbf{Nelly}'s weather readings, \textbf{Flic}'s cats, \textbf{Marc}'s gif-based career advice, \textbf{Conor}'s machine learning rants, \textbf{Ivana}'s coffee breaks, and \textbf{James}'s piano. Thank you for being there through the worst of it all.

This work was supported in part by the EPSRC Centre for Doctoral Training in Data Science, funded by the UK Engineering and Physical Sciences Research Council (grant EP/L016427/1) and the University of Edinburgh. In addition, part of the work was done while interning at Google. Thank you for making it possible for me to pursue a PhD by funding this research. 
\end{acknowledgements}

\standarddeclaration

\includepdf{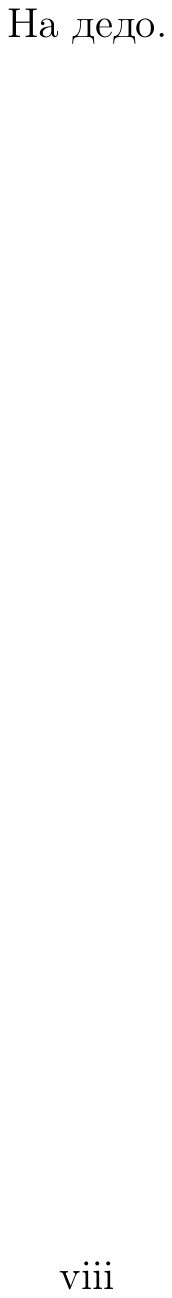}

\tableofcontents


\end{preliminary}


\include{chapters/intro}

\include{chapters/background}

\cleardoublepage
\part{Static optimisation for probabilistic programming} \label{part:static}
\include{chapters/part1-intro}

\include{chapters/pl-background}

\include{chapters/slicstan1}

\include{chapters/slicstan2}

\cleardoublepage
\part{Dynamic optimisation for probabilistic programming} \label{part:dynamic}
\include{chapters/part2-intro}
\include{chapters/autoreparam}
\include{chapters/conclusion}

\appendix

\bibliographystyle{ACM-Reference-Format}


{\footnotesize \bibliography{thesis}}

\end{document}

%% file: chapters/intro.tex
\cleardoublepage
\chapter{Introduction}

\textit{Bayesian statistics} provides a set of rigorous techniques that allow us to analyse and interpret data, and to make predictions while carefully quantifying uncertainty. 
It has been an invaluable tool in many areas where it is important to incorporate domain knowledge in the modelling of the data, or where data is scarce. 
In Bayesian terms, we define a \textit{statistical model} that encodes our assumptions about an underlying process based on some \textit{unknown variables} and some real-world data that we \textit{observe}. Through \textit{Bayesian inference} (which I will also refer to as just \textit{inference} in this dissertation) we can obtain the conditional distribution of the unknown quantities given the data, and compute expectations under it, while rigorously keeping track of uncertainties. 

But inference is not an easy task. Performing inference correctly, reliably and efficiently, requires substantial knowledge of probability and statistics, it is an active area of research, and it can be time-consuming even for experts. 
Thus, it can be difficult for non-statistical fields to adopt Bayesian statistics, despite the benefits it has to offer. 

\textit{Probabilistic programming} aims to decouple modelling from inference. It promises to allow modellers to work close to their domain of expertise, without having to implement inference algorithms from scratch. The idea of probabilistic programming languages is to specify a model in the form of a program, which typically (but not always) encodes the way in which the data is believed to have been generated. This program is then automatically or semi-automatically compiled to an inference algorithm, which can compute expectations of interest without users having to implement computational details. 

While probabilistic programming has significantly grown as an area in the past decade, its dream to democratise Bayesian statistics is still some way ahead. 
Different inference algorithms come with different constraints, advantages, and disadvantages. General algorithms, which work with a wide range of models, typically take longer to give a reliable result. Tailoring an algorithm to a specific problem can result in a considerably more efficient inference, but using the same algorithm for a different model might not be possible. 
Probabilistic programming languages face this same problem and are often forced into a trade-off between generality of the modelling language and efficiency of inference.

The focus of this dissertation is on program analysis techniques for probabilistic programming. Many probabilistic programming languages leverage the formal foundations on which they have been built to compile to a particular algorithm or perform symbolic inference. But whether the underlying program structure can be utilised to improve inference in a program-specific manner has been largely unexplored. 
The main question of this dissertation is:
\textit{Can program analysis improve inference in probabilistic programming?}
I claim the answer to this question is \textit{yes} and I show three ways in which program analysis of probabilistic programs can support automatic inference.

\section*{Contributions and outline of the dissertation}

The dissertation is in two parts, which focus on static and dynamic analysis respectively.
\autoref{ch:background} provides background and motivation relevant to both parts. It gives a short introduction to Bayesian inference, covering different approaches and discussing their advantages and disadvantages. It also gives an overview of probabilistic programming languages, dividing them in three categories, and listing some of their strengths and constraints.
The brief~\autoref{ch:pl-background}, on the other hand, gives background on formal treatment of programming languages that is relevant to \autoref{part:static} only.

Chapters \ref{ch:slicstan}--\ref{ch:autoreparam} each focus on one of the \textit{three contributions} of this dissertation:
\begin{enumerate}
\item \autoref{ch:slicstan} describes the first formal treatment of the probabilistic programming language Stan, and a semantics-preserving procedure that allows for a more compositional, self-optimising version of Stan, called SlicStan. \autoref{ch:slicstan} is based on the following publication:
\begin{quote}
	Maria I Gorinova, Andrew D Gordon, and Charles Sutton. Probabilistic programming with densities in SlicStan: Efficient, flexible, and deterministic. \textit{Proceedings of the ACM on Programming Languages} 3, Issue POPL, Article 35 (January 2019).
\end{quote}

\item \autoref{ch:slicstan2} extends Stan/SlicStan with explicit support for discrete parameters, through a semantics-preserving transformation that firstly marginalises out discrete variables from the program and then re-draws them. This can be seen as an automatic program-specific composition of two inference algorithms.
\autoref{ch:slicstan2} is based on the following publication: 
\begin{quote}
	Maria I Gorinova, Andrew D Gordon, Charles Sutton, and Matthijs V\'ak\'ar. \hspace{-4pt} Conditional independence by typing.\hspace{-3pt} \textit{ACM Transactions on Programming Languages and Systems} 44, Issue 1, Article 4 (March 2022).
\end{quote}

\item \autoref{ch:autoreparam} gives a way to automatically reparameterise probabilistic models through effect-handlers and shows how the process of finding a suitable parameterisation of a probabilistic model can be automated through a variational inference pre-processing. 
The chapter is based on the following publication: 
\begin{quote}
	Maria I Gorinova, Dave Moore, and Matthew D Hoffman. Automatic Reparameterisation of Probabilistic Programs. \textit{Proceedings of the 37th International Conference on Machine Learning} (July 2020).
\end{quote}
\end{enumerate}

%% file: chapters/background.tex
\cleardoublepage
\Chapter{Probabilistic programming}{Inference and languages} \label{ch:background}

What is probabilistic programming and why is it difficult? This chapter gives a brief overview of the topic of probabilistic programming, motivating the work of this thesis and introducing some preliminaries. 
It gives a gentle introduction to the idea of probabilistic programming (\autoref{sec:bg-simplePP}), and discusses the process and challenges of 
Bayesian inference and of probabilistic modelling itself (\autoref{sec:bg-inference}). Finally, the chapter gives background on some of the most well-known current probabilistic programming languages, analysing their advantages, shortcomings, and constraints (\autoref{sec:bg-ppls}).

\section{A very simple probabilistic program} \label{sec:bg-simplePP}

Suppose Anna flips two fair coins, observes the result, and tells Ben that they are \textit{not} both heads. How does Ben's belief about the flip result of each coin changes based on this new information? 

Anna's flips generate one of four possible scenarios: both coins are heads, both are tails, the first is heads and the second is tails, or the first is tails and the second is heads. But one of these scenarios becomes impossible given the extra information Anna gives us. Each of the two coins is heads in only one of the remaining three equally likely scenarios. Thus, the probability that, say, the first coin is heads given they are not both heads is $1/3$. 

We can express such questions about the uncertainty over phenomena given observations using probabilistic programming languages (PPLs). In general, such languages are characterised by \textit{two} facilities that are not usually present in conventional programming. These are the ability to specify \textit{random variables} and the ability to \textit{observe data}. 

For example, we can express Anna and Ben's problem as a probabilistic program: 
\begin{lstlisting}[numbers=left, stepnumber=1,numbersep=-4pt,numberstyle=\scriptsize\color{seabornblue}]
		c1 ~ bernoulli(0.5)
		c2 ~ bernoulli(0.5)
		bothHeads = c1 and c2
		observe(not bothHeads)
\end{lstlisting}

Each coin flip is a \textit{Bernoulli} random variable, which is either 1 (for heads) or 0 (for tails) with probability $0.5$. In the program above, lines 1 and 2 define these two variables, naming them $c_1$ and $c_2$. Line 3 introduces a third random variable, $\mathrm{bothHeads}$, which is the logical `and' between $c_1$ and $c_2$: it is 1 only when both coin flips result in heads, and 0 otherwise. 
Finally, line 4 \emph{observes} the \textit{data}: the coin flips did not both result in heads.  
A PPL that can interpret this program will give us the conditional probability distribution of the \textit{unobserved} variables in the program given the data. In our case, that is:
$$p(c_1, c_2 \mid \kw{not}~\mathrm{bothHeads}) = \begin{cases}
0 &\text{if } c_1 = c_2 = 1 \text{ or } c_1, c_2 \notin \{0, 1\}^2 \\
1/3 &\text{otherwise.}  
\end{cases}$$

\paragraph{Notation and terminology} More concretely, a probabilistic program defines a joint density over some random variables. We typically divide these variables into \textit{parameters} $\params$ and \textit{data} $\data$ and write $p(\params, \data)$ for their joint distribution, which is also the product of the \textit{prior} on the parameters $p(\params)$ and the \textit{likelihood} of the data $p(\data \mid \params)$.
We are interested in the \textit{posterior} distribution
\begin{equation} \label{eq:bayes}
p(\params \mid \data) = \frac{p(\params)p(\data \mid \params)}{p(\data)}. 
\end{equation}
Specifically, we are interested in extracting information from this distribution, in order to answer different queries. This usually means computing \textit{expectations} under this posterior distribution, for example the expectation of a function $f(\params)$:%
\footnote{This expression assumes continuous parameters $\params$. For mixed discrete and continuous parameters, we can replace the integration with an appropriate combination of integration and summation. Generally, integration is used as a shorthand for such integration-summation combination throughout this thesis.}
\begin{equation} \label{eq:expectation}
\expectation[p(\params \mid \data)]{f(\params)} = \int f(\params)p(\params \mid \data) \dif{\params}.
\end{equation}
The process of deriving or approximating the posterior distribution, in a form that allows us to compute expectations under it, is what we call \textit{Bayesian inference}.%
\footnote{This is only one way of defining ``inference''. Statistical inference can more generally be understood as deriving properties (expectations, but also confidence / credible intervals) of some underlying distribution.}

But why do we need a dedicated language to do any of this? Can we not simply use Equations~\ref{eq:bayes} and~\ref{eq:expectation} to compute any expectation we like? The next sections describe some of the challenges of performing Bayesian inference in practice and, in turn, the challenges faced by PPLs: the languages that aim to democratise this difficult process. 

\newpage
\section{Bayesian inference} \label{sec:bg-inference}

\subsection{Deriving a solution analytically}

The first question we might ask is can we not perform inference analytically. Indeed, given the joint $p(\params, \data)$, \autoref{eq:bayes} gives us the posterior $p(\params \mid \data) = \frac{p(\params, \data)}{p(\data)} = \frac{p(\params, \data)}{\int p(\params, \data) \dif \params}$. But the \textit{normalising constant} (also referred to as the \textit{marginal likelihood} or \textit{evidence}) $Z(\data) = \int p(\params, \data) \dif \params$ is not always available in a closed form and it is infeasible to compute in the general case. 
Suppose, for example, that we are working with $N$ Bernoulli parameters $\params = \left(\theta_1, \dots, \theta_N\right)$. Then $Z(\data) = \sum_{\hat{\theta_1}=0}^{1} \dots \sum_{\hat{\theta_N}=0}^{1} p(\theta_1 = \hat{\theta_1}, \dots, \theta_N = \hat{\theta_N}, \data) = \sum_{\params \in \{0,1\}^N}p(\params, \data)$. The complexity of computing this expression is $O(2^N)$, which becomes infeasible for larger $N$.

What is worse is that even if we assume an oracle that gives us $Z(\data)$, computing expectations under $p(\params \mid \data)$ is still infeasible: $\expectation[p(\params \mid \data)]{f(\params)} = \sum_{\params \in \{0,1\}^N}p(\params, \data)f(\params)$. In practice, such expectations are typically approximated. One way to approximate an expectation under some $p(\mathbf{x})$ is through \textit{Monte Carlo estimation} \cite{MonteCarlo}:
\begin{equation} \label{eq:montecarlo}
	\expectation[p(\mathbf{x})]{f(\mathbf{x})} \approx \frac{1}{I} \sum_{i=1}^{I}f(\hat{\mathbf{x}}^{(i)}) \qquad\qquad \text{where}~\hat{\mathbf{x}}^{(i)} \sim p(\mathbf{x}) 
\end{equation}

In other words, if we can obtain a set of samples $\hat{\mathbf{x}}^{(i)}$ from the distribution of interest $p(\mathbf{x})$, then we can use this set to estimate the expectation of any function under $p(\mathbf{x})$.

\subsection{Rejection sampling and the curse of dimensionality} \label{ssec:curse}

This leads us to a second idea of how to perform inference: can we not simply sample $(\hat{\params}^{(i)}, \hat{\data}^{(i)})$ from the joint $p(\params, \data)$, discard any samples where $\hat{\data}^{(i)}$ does not match the actual observed data, and use the remaining samples for Monte Carlo estimation? 

For instance, in the coins example, we can generate $S = \{(c_1^{(i)}, c_2^{(i)}, \mathrm{bothHeads}^{(i)})\}_{i=1}^I$, where $c_1^{(i)} \sim \mathrm{bernoulli}(0.5)$, $c_2^{(i)} \sim \mathrm{bernoulli}(0.5)$, and $\mathrm{bothHeads}^{(i)} = c_1^{(i)} \wedge c_2^{(i)}$. \textit{Rejecting} all samples such that $\mathrm{bothHeads}^{(i)} = 1$, leaving us with $S' = \{ (c_1^{(i)}, c_2^{(i)}, \mathrm{bothHeads}^{(i)}) \in S \mid \mathrm{bothHeads}^{(i)} = 0 \}$. These remaining samples $S'$ are samples from the posterior distribution $p(c_1, c_2 \mid \mathrm{bothHeads} = 0)$. 

This is a particular instantiation of the \textit{rejection sampling algorithm}, and is a valid inference algorithm, as (after scaling) $p(\params \mid \data)$ is entirely contained by $p(\params, \data)$:
\begin{equation}
p(\hat{\params}, \data) = p(\data)p(\hat{\params} \mid \data) \geq C p(\params \mid \data)
\quad \text{for all } \hat{\params} \text{ and } C \text{ a constant}
\end{equation}

\begin{figure}
	\centering
	\begin{subfigure}{0.32\textwidth}
		\includegraphics[width=\textwidth]{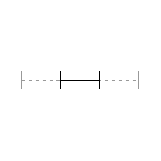}
	\end{subfigure}
	\begin{subfigure}{0.32\textwidth}
		\includegraphics[width=\textwidth]{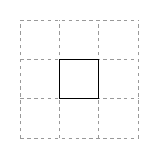}
	\end{subfigure}
	\begin{subfigure}{0.32\textwidth}
		\includegraphics[width=\textwidth]{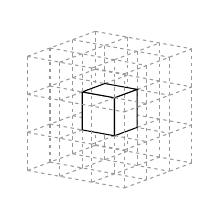}
	\end{subfigure}
	\caption{Illustration of the curse of dimensionality: the ratio between some volume of interest and a volume that contains it, diminishes as dimensionality increases. The figure is taken from Michael Betancourt's tutorial on \textit{Probabilistic Computation} \url{https://betanalpha.github.io/assets/case_studies/probabilistic_computation.html}.}
	\label{fig:volume}
\end{figure}

While simple, general, and exact,  this algorithm does not scale. To see why, consider continuous parameters $\params$ and suppose that the prior over $\params$ is an $N$-dimensional uniform distribution between $0$ and $1$. Suppose also, that the data is such that the posterior of $\params$ is an $N$-dimensional uniform distribution between $1/3$ and $2/3$. The prior forms an N-dimensional hypercube of side $1$, and the posterior is an $N$-dimensional hypercube of side $1/3$ centred at the middle of the prior hypercube. \autoref{fig:volume} shows this for $N \in \{1,2,3\}$. Our rejection sampling algorithm would choose at random a point inside of the big hypercube and only accept it if it is also inside the smaller hypercube. The acceptance rate $R$ is proportionate to the ratio between volumes of the two hypercubes:
\begin{equation} \label{eq:volume-ratio}
	R = \frac{(1/3)^N}{1^N} = \frac{1}{3^N} 
\end{equation}
Even if we are to sample from a prior closer to the posterior, say uniform between $1/3-\epsilon$ and $2/3+\epsilon$, for any small $\epsilon$, the ratio between volumes, and thus the acceptance rate, is:
\begin{equation} \label{eq:volume-ratio2}
R_{\epsilon} = \frac{(1/3)^N}{(1/3 + 2\epsilon)^N} = \frac{1}{(1 + 6\epsilon)^N} 
\end{equation}
In either case, as the dimensionality increases, the volume outside of the inner hypercube increases exponentially, and the acceptance rate approaches $0$.
This is a particular instantiation of the \textit{curse of dimensionality} \cite[section 1.4]{Bellman1961,Bishop}. 

The curse of dimensionality is why many numerical and sampling methods, such as numerical integration, rejection sampling, and importance sampling, are impractical for problems with more than few dimensions. It also motivates the search for more effective inference solutions, which exploit various properties of probability distributions, such as smoothness or conditional independencies.

\subsection{Markov chain Monte Carlo inference}

Markov chain Monte Carlo (MCMC) is a class of sampling algorithms, introduced by \citet{MCMCFirst} and reviewed by \citet{MCMC} and \citet{MCMCIain}, which perform sampling by transitioning according to a Markov chain. This chain is carefully designed, so that its \textit{stationary distribution} is the target probability distribution (for example the posterior distribution in the case of inference). Drawing samples according to the Markov chain eventually converges to drawing samples from the target, which makes MCMC methods \textit{asymptotically exact}: they are exact in the presence of infinite amount of samples.

Consider the problem of sampling from some $p(\params)$. Let $\pi(\params \mid \params')$ be the transition probabilities (also referred to as \textit{kernel}) of a homogeneous Markov chain. 
Given some initial sample $\params^{(0)}$, we can generate a sequence of samples 
$\params^{(1)} \sim \pi(\params^{(1)} \mid \params^{(0)})$, 
$\params^{(2)} \sim \pi(\params^{(2)} \mid \params^{(1)})$, $\dots$,
$\params^{(N)} \sim \pi(\params^{(N)} \mid \params^{(N-1)})$. 
This sequence converges towards samples from $p(\params)$, provided:
\begin{enumerate}
	\item the chain is \textit{ergodic}, 
	meaning it is aperiodic and it is possible to ``reach'' any state $\params'$ from any other state $\params$ following a finite sequence steps of non-zero probability.
	Then, no matter the value of the initial $\params^{(0)}$, $\pi(\params^{(N)} \mid \params^{(0)})$ reaches a unique \textit{stationary distribution} (also called \textit{equilibrium distribution }) $\pi_{\infty}(\params^{(N)})$ in the limit of $N \rightarrow \infty$:
	\begin{equation} \label{eq:ergodic}
		\lim_{N \rightarrow \infty} \pi(\params^{(N)} \mid \params^{(0)}) = \pi_{\infty}(\params^{(N)})
	\end{equation}
	\begin{equation} \label{eq:stationary}
		\pi_{\infty}(\params) = \int \pi(\params \mid \params') \pi_{\infty}(\params') \dif \params'
	\end{equation}
	
	\item $p(\params)$ is this stationary distribution: $\pi_{\infty}(\params) = p(\params)$.
\end{enumerate}

As long as these conditions are met, drawing samples, one by one, from $\pi(\params^{(n)} \mid \params^{(n-1)})$ is a valid sampling algorithm. When used to sample from a posterior distribution, it is also a valid inference strategy. 
While MCMC guarantees exact inference in the presence of infinitely many samples, the quality of inference in practice is hugely influenced by the shape of the underlying distribution and the choice of kernel $\pi(\params \mid \params')$.

The MCMC family of sampling algorithms is very large, and a full review is out of scope for this thesis, but is studied in detail by many probabilistic reasoning textbooks
(\citealt[Chapters~29 and 30]{MacKay}; \citealt[Chapter 11]{Bishop}; \citealt[Chapter 27]{Barber}; \citealt[Chapter 24]{Murphy}).
In the rest of this section, I will focus on one particular MCMC algorithm, which is closely related to the work of this thesis: Hamiltonian Monte Carlo.

\subsubsection*{Hamiltonian Monte Carlo}

Hamiltonian Monte Carlo (HMC) is an MCMC algorithm introduced by \citet{HMCfirst} and popularised by \citet{HMCfirstNeals}. It makes use of the gradient of the target density function as a way to efficiently explore the underlying probability distribution. HMC is typically applied to problems where the number of parameters in the model is known and fixed, although extensions that work for nonparametric models have also been recently introduced \cite{mak2021nonparametric}. 

In brief, HMC works by translating the problem of sampling from a probability distribution to exploring the dynamics of a particle in a Hamiltonian system.
Suppose that the target density $p(\params)$ is differentiable almost everywhere. Suppose also that we are given a function $p^*(\params)$ that is proportionate to the $p(\params)$; that is $p(\params) = \frac{p^*(\params)}{Z}$ for the normalising constant $Z = \int p^*(\params) \dif \params$.
Now consider the multidimensional surface given by $-\log p^*(\params)$. A particle on that surface can be described by its position $\params$ and momentum $\mathbf{p}$. Its \textit{potential energy} $E(\params)$ is given by the height of the surface at $\params$ and its \textit{kinetic energy} $K(\mathbf{p})$ is given by the magnitude of the momentum $\mathbf{p}$ and the mass $m$ of the particle:

$$E(\params) = - \log p(\params) - \log Z \qquad\qquad\qquad K(\mathbf{p}) = \frac{1}{2m}\norm{\mathbf{p}}^2$$              

\begin{figure}
	\centering
	\includegraphics[width=\textwidth]{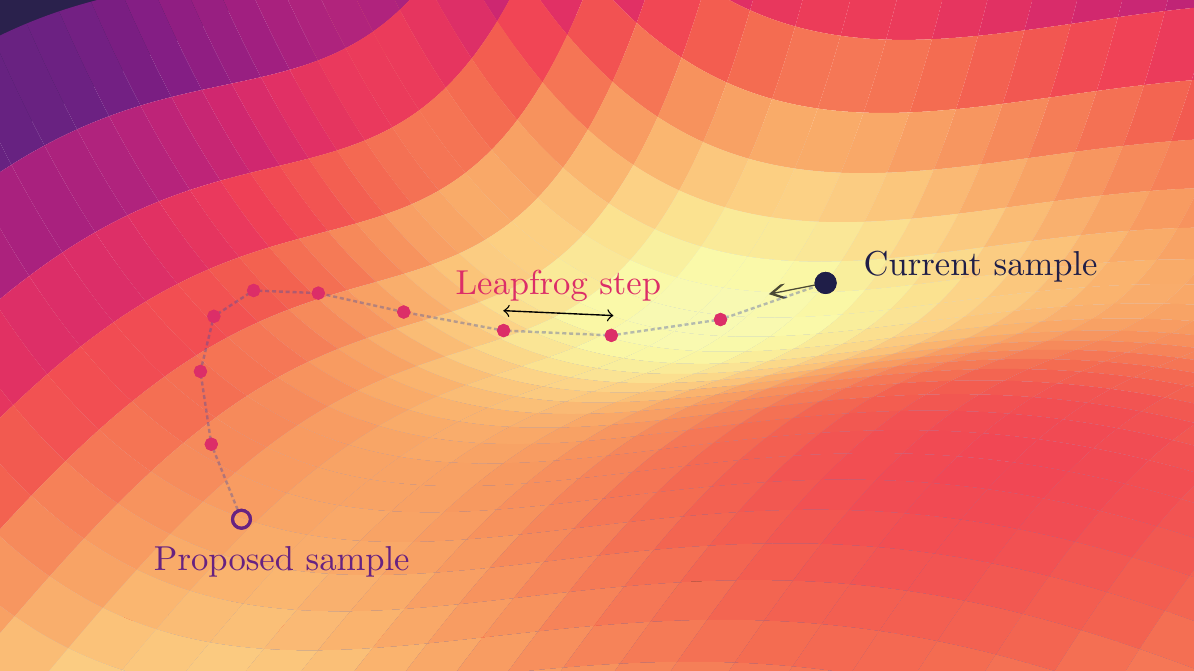}
	\caption{An illustration of a single HMC iteration in a 2-dimensional parameter space. The current sample is given an initial momentum at random; we take 10 leapfrog steps to simulate the particle trajectory and obtain a candidate sample.}
	\label{fig:hmc}
\end{figure}

In a system with a single particle and no friction, no energy is lost and the
Hamiltonian, the sum of potential and kinetic energy, stays constant:
$$H(\params, \mathbf{p}) = E(\params) + K(\mathbf{p})$$

The change of the position and momentum over time $t$ is described by a system of differential equations:
$$ \frac{\dif \params}{\dif t} = \nabla K(\mathbf{p}) = \frac{1}{m} \mathbf{p}  \qquad\qquad\qquad \frac{\dif \mathbf{p}}{\dif t} = - \nabla E(\params) $$

Here $\nabla E(\params) = \begin{pmatrix} \frac{\partial E(\params)}{\partial \theta_1} & \dots & \frac{\partial E(\params)}{\partial \theta_N} \end{pmatrix}$ and $\nabla K(\mathbf{p}) = \begin{pmatrix} \frac{\partial K(\mathbf{p})}{\partial p_1} & \dots & \frac{\partial K(\mathbf{p})}{\partial p_N} \end{pmatrix}$ denote the gradients of $E(\params)$ and $K(\mathbf{p})$ respectively.

Equipped with these equations, we can simulate (run forwards in time) the ``physical'' system for any initial position $\params$ and momentum $\mathbf{p}$. Now consider the MCMC transition kernel $\pi(\params, \mathbf{p}, t \mid \params', \mathbf{p}', t')$, which is $1$ if simulating the system between time $t'$ and $t$, starting at an initial position $\params'$ and initial momentum $\mathbf{p}'$, results in position $\params$ and momentum $\mathbf{p}$, and it is $0$ otherwise. As the system is deterministic, we can also recover the initial position and momentum by simulating back in time starting from the final position and momentum.%
\footnote{This is equivalent to simulating forwards in time starting from the final position $\params'$ and the reversed final momentum $-\mathbf{p}'$.}%
That is:
\begin{equation}
\pi(\params, \mathbf{p}, t \mid \params', \mathbf{p}', t') = \pi(\params', \mathbf{p}', t' \mid \params, \mathbf{p}, t)
\end{equation}

Consider the joint density $p_H(\params, \mathbf{p}) = \frac{\exp(- H(\params, \mathbf{p}))}{Z_H}$, where $Z_H = \int \exp(- H(\params, \mathbf{p})) \dif \params \dif \v{p} = Z \int \exp(-K(\mathbf{p})) \dif \v{p}$ is a normalising constant.
As the Hamiltonian stays constant, we have $p_H(\params, \mathbf{p}) = p_H(\params', \mathbf{p}')$, giving us:
\begin{equation} \label{eq:detailed_balance}
	p_H(\params, \mathbf{p})\pi(\params, \mathbf{p}, t \mid \params', \mathbf{p}', t') = p_H(\params', \mathbf{p}')\pi(\params', \mathbf{p}', t' \mid \params, \mathbf{p}, t)
\end{equation}

\autoref{eq:detailed_balance} is known as \textit{detailed balance} and it is a sufficient condition for showing that $p_H(\params, \mathbf{p})$ is the stationary distribution of $\pi$ \cite[Theorem~17.2.3]{Murphy}. Thus, simulating the sliding particle physical system produces samples from the joint density  $p_H(\params, \mathbf{p})$. Marginalising over $\mathbf{p}$ gives us:
\begin{align*}
	\int_{-\infty}^{+\infty}p_H(\params, \v{p}) \dif\v{p} &= \int_{-\infty}^{+\infty}\frac{1}{Z_H}\exp(- E(\params))\exp(-K(\v{p})) \dif\v{p} \\
	&= \frac{1}{Z}\exp(- E(\params)) \int_{-\infty}^{+\infty}\frac{Z}{Z_H}\exp(-K(\v{p})) \dif\mathbf{p} \\
	&= p(\params)
\end{align*}

In other words, we can obtain samples from $p(\params)$ by augmenting the parameter space with the momentum variables $\v{p}$, running a simulation of the Hamiltonian system described above, and discarding the samples obtained for $\v{p}$. In practice, we need to discretise time in order to run the simulation, which in HMC is usually done using \textit{leapfrog integration}. The error introduced by discretising the simulation is corrected with an accept/reject step similar to the Metropolis-Hastings algorithm \cite{MH}.

A single iteration of HMC is visually described in \autoref{fig:hmc}. In brief, obtaining a new sample $\params_{n+1}$ given a current sample $\params_{n}$ is done as follows:
\begin{enumerate}
\item Sample an initial momentum $\v{p} \sim \normal(\v{0}, mI)$, where $I$ is the identity matrix, and compute the Hamiltonian of the system $H = E(\params_{n}) + K(\v{p})$.
\item Simulate the system starting at position $\params_n$ and momentum $\v{p}$. This step requires $E(\params)$ to be differentiable with respect to $\params$ and uses the gradient $\nabla E(\params)$ to make steps (some small $\epsilon$ in size) in the direction of steepest descent. After $L$ leapfrog steps the particle is at position $\params^*$ and has momentum $\v{p}^*$.
This final position of the particle, $\params^*$, is the proposed sample. 
\item Decide whether to accept ($\params_{n+1} = \params^*$) or reject ($\params_{n+1} = \params_n$) the candidate sample. This is done based on the difference between the re-evaluated Hamilton 
$H^* = E(\params^*) + K(\v{p}^*)$ and the initial Hamilton $H$. The bigger this difference, the less accurate was the discretised simulation at approximating the behaviour of physical system, and the less likely it is to accept the candidate sample. When the difference is very big, we call the sample step \textit{divergent}. Divergences can indicate problems with inference and are an important diagnostics tool, as shown next and in \autoref{fig:neals_all}.
\end{enumerate}

\subsubsection*{Advantages and shortcomings}

MCMC methods are asymptotically exact: for an infinite number of samples, 
sampling from the transition distribution corresponds to sampling from $p(\params)$ 
exactly. In practice, this means that to obtain a reliable approximation of a distribution, we need to sample for a long time, until the Markov chain \textit{converges}. Unfortunately, there is no way to calculate in advance for how long we should run a given chain to achieve some desired accuracy \cite{Tierney1994Convergence, Brooks1998Convergence}.
In particular, the geometry of the target distribution has a dramatic effect on the performance of MCMC. \textit{Well-conditioned} distributions that have a small ratio between the variable with highest variance and the variable with the lowest variance, will typically be easier to sample from than distributions where this ratio is large. This leads to some pathological cases, where no number of samples smaller than infinity provides a correct approximation of the distribution of interest.

\begin{figure}
\begin{subfigure}{0.49\textwidth}
	\includegraphics[width=\textwidth]{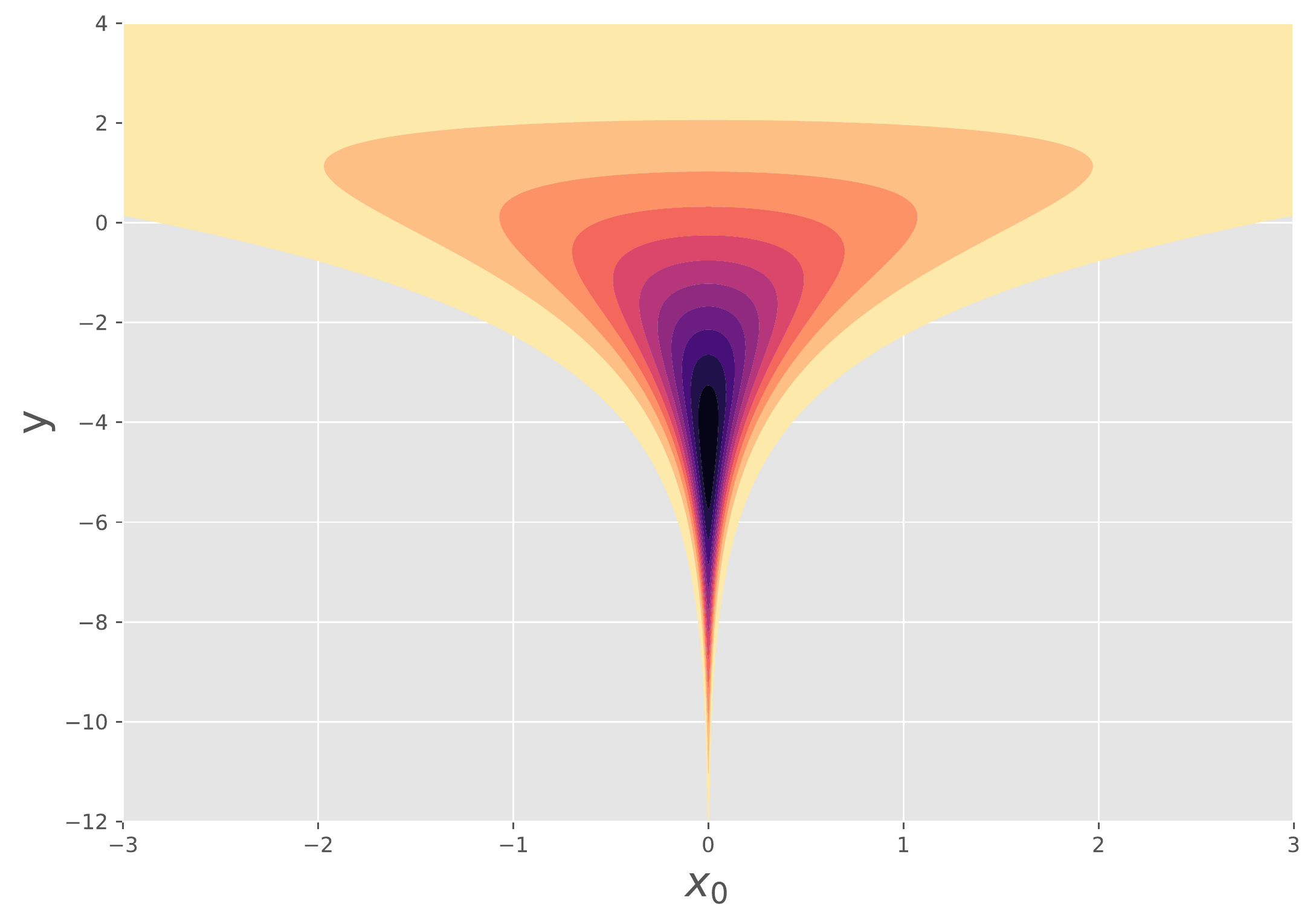}
	\caption{Neal's funnel density.}
	\label{fig:neals}
\end{subfigure}
\begin{subfigure}{0.49\textwidth}
	\includegraphics[width=\textwidth]{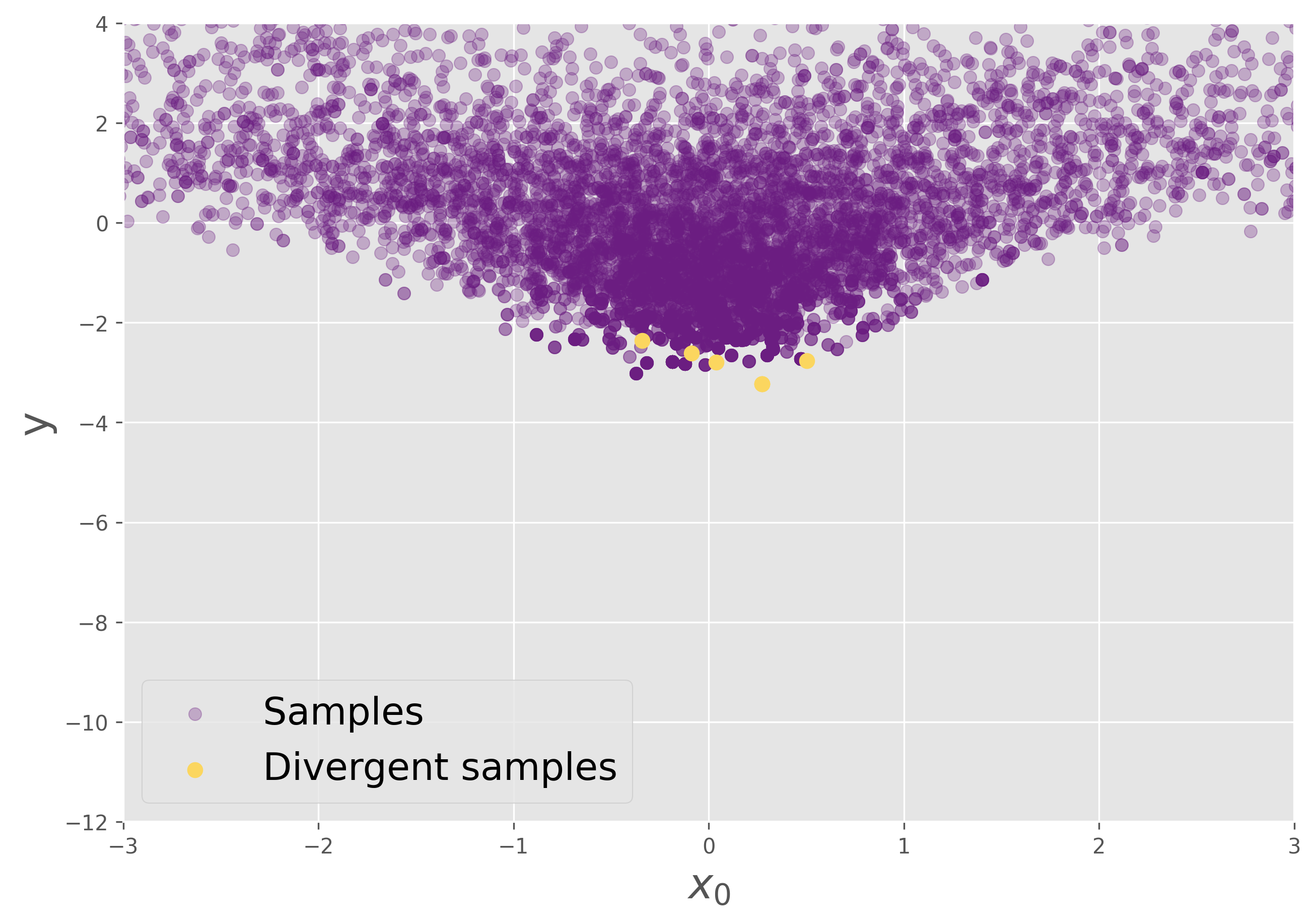}
	\caption{$10000$ samples drawn with HMC.}
	\label{fig:neals_scatter}
\end{subfigure}
\caption{Neal's funnel.}
\label{fig:neals_all}
\end{figure}

For example, consider \textit{Neal's funnel}, which is a model chosen by \citet{SliceSampling} to demonstrate the difficulties Metropolis--Hastings \cite{MH} runs into when sampling from a distribution with strong non-linear dependencies. The model defines a density over variables $\mathbf{x}$ and $y$:
$$y \sim \normal(0,3) \qquad x_i \sim \normal(0, \exp(y / {2})) \qquad \text{for } i = 1,\dots,9$$
The density has the form of a 10-dimensional funnel (thus the name \textit{``Neal's funnel''}), with a very sharp neck, as shown in \autoref{fig:neals}.
MCMC methods, including HMC, have trouble obtaining samples from the neck of the funnel, because there exists a strong non-linear dependency between $\mathbf{x}$ and $y$, and the posterior geometry is difficult for the sampler to explore well. \autoref{fig:neals_scatter} shows $10 000$ samples drawn from Neal's funnel using HMC. Despite half of the probability mass being inside of the funnel, HMC is not able to explore it fully due to its high curvature. Samples where a divergence occurred are shown in yellow, concentrating around the region that HMC's leapfrog integrator finds problematic.

Neal's funnel is a typical example of the dependencies that can occur in hierarchical models in practice, and therefore it highlights the importance of MCMC diagnostic techniques that can detect possible inference problems. Many such techniques exist, including examining HMC divergences, comparing the within and between variance of several different MCMC chains, prior predictive and posterior predictive checks. In addition, there are various ways to address inference problems, one of which, model reparameterisation, I discuss in detail in \autoref{ch:autoreparam}. 
For an excellent introduction to various diagnostic techniques and ways to address computational problems, refer to \citet{BayesianWorkflow}.

\subsection{Variational inference} \label{ssec:vi}

In practice, sampling methods can be very computationally demanding, as they sometimes require a huge number of iterations to obtain a reasonable approximation for a target distribution. 
Variational inference (VI) methods (\citealt[Chapter~33]{MacKay}; \citealt[Chapter 10]{Bishop}; \citealt[Chapter 21]{Murphy}; \citealt{Blei2017Variational}) take a different approach that trades asymptotic guarantees for efficiency. The main idea of variational inference is to use \textit{optimisation} instead of sampling, to obtain an analytical approximation to a distribution of interest.
In brief, given a tractable family of distributions, VI tries to find a distribution from that family that best approximates the target distribution. 

Consider some \textit{family} $Q$ of distributions over the parameters $\params$. 
We call variational inference the process of finding some optimal $q^*(\params) \in Q$, such that $q^*(\params)$ is as close as possible to the distribution of interest, for example, the posterior $p(\params \mid \data)$:
\begin{equation} \label{eq:vi}
	q^*(\params) = \argmin{q(\params) \in Q}{\mathrm{d}(p(\params \mid \data), q(\params))} 
\end{equation}
Here, $\mathrm{d}((p(\params \mid \data), q(\params)))$ encodes the \textit{dissimilarity} between $p$ and $q$. Different dissimilarity metrics can be used, but the one that is most common in variational inference is the \textit{Kullback–Leibler (KL) divergence} \cite{KL} from $q$ to $p$: 
\begin{equation} \label{eq:kl}
	\kl{q(\params)}{p(\params \mid \data)} = \expectation[q(\params)]{\log q(\params) - \log p(\params \mid \data)}
\end{equation}
However, we cannot directly use this metric in an objective, as we do not know $p(\params \mid \data)$. Usually, the KL divergence is used to derive a lower bound for the evidence $p(\data)$. Note that $\kl{q(\params)}{p(\params \mid \data)} \geq 0$, and only equals $0$ when $p$ and $q$ are identical. 
This results is knows as \textit{Gibbs' inequality} \cite[Section~2.6]{MacKay}.
We derive:
\begin{align*}
\kl{q(\params)}{p(\params \mid \data)} 
&= \expectation[q(\params)]{\log q(\params) - \log p(\params \mid \data)} \\
&= \expectation[q(\params)]{\log q(\params)} - \expectation[q(\params)]{\log \frac {p(\params, \data)}{p(\data)}}\\
&= \expectation[q(\params)]{\log q(\params)} - \expectation[q(\params)]{\log p(\params, \data)} + \expectation[q(\params)]{\log p(\data)} \\
&= \expectation[q(\params)]{\log q(\params)} - \expectation[q(\params)]{\log p(\params, \data)} + \log p(\data)\\
&\geq 0
\end{align*}
And define the \textit{evidence lower bound (ELBO)} as:
\begin{equation} \label{eq:elbo}
	\mathrm{ELBO} =
	\expectation[q(\params)]{\log p(\params, \data)} - \expectation[q(\params)]{\log q(\params)} 
	\leq \log p(\data)
\end{equation}
As $\log p(\data)$ is a constant with respect to $\params$, minimising the KL-divergence \eqref{eq:kl} is equivalent to maximising the ELBO \eqref{eq:elbo}. Assuming that we are given $p(\params, \data)$ (or, in the more general case, any function proportionate to $p(\params \mid \data)$), and that we can efficiently estimate expectations under $q(\params)$, we can re-define our variational objective as maximising the expectation lower bound:
\begin{equation} \label{eq:vi-elbo}
	q^*(\params) = \argmax{q(\params) \in Q}{\left[
		\expectation[q(\params)]{\log p(\params, \data)} - \expectation[q(\params)]{\log q(\params)}\right]} 
\end{equation}

\begin{figure}[!h]
	\includegraphics[width=\textwidth]{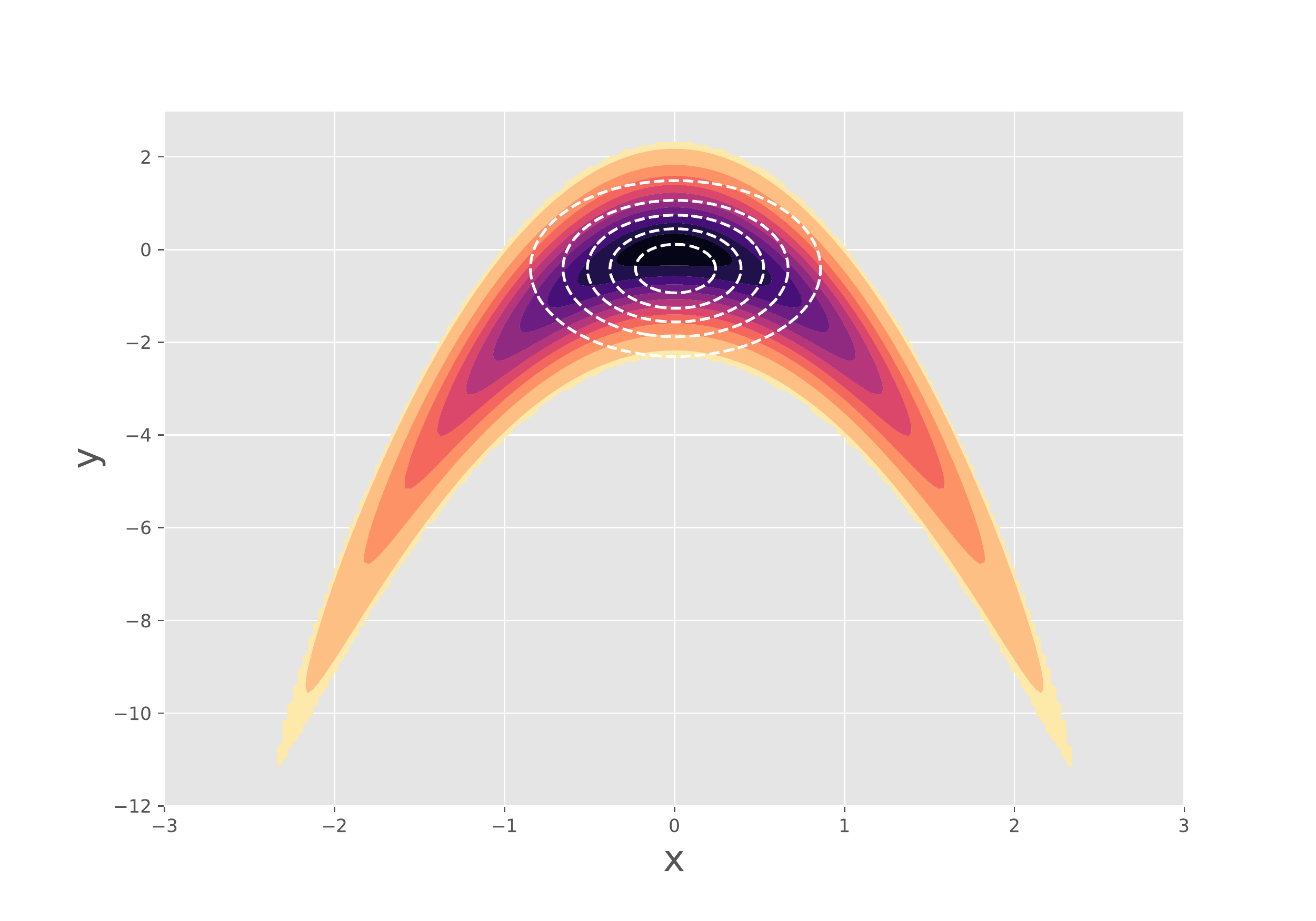}
	\caption{An example of the variational family of choice not capturing the complexity of the target distribution. The target banana-shaped distribution is given by $x \sim \normal(0, 1)$ and $y \sim \normal(-2x^2, 1)$. The dashed white lines show the mean-field variational posterior.}
	\label{fig:vi-example}
\end{figure}

There are different ways to choose a variational family $Q$, so that expectations under distributions in the family are easy to compute (\citealt[Chapter 10]{Bishop}, \citealt[Chapter 21]{Murphy}). One way that is increasingly used, and especially in the context of probabilistic programming, is to parameterise the distribution using some set of \textit{variational parameters} $\phi$: a distribution $q(\params; \phi) \in Q$ is entirely determined by its variational parameters $\phi$. In addition, the family $Q$ is chosen so that distributions belonging to it are easy to sample from. This allows us to approximate the expectations in the ELBO by Monte Carlo averaging. \autoref{eq:vi-elbo} becomes:
\begin{equation} \label{eq:vi-parameterised-elbo}
	\phi^* = \argmax{\phi}{
		\sum_{\hat{\params} \sim q(\params; \phi)}\left[\log p(\hat{\params}, \data) - \log q(\hat{\params}; \phi)\right]} 
\end{equation}

Many modern differentiable programming frameworks, such as TensorFlow \cite{Tensorflow} and PyTorch \cite{PyTorch}, can automatically differentiate through such expressions. Thus, we can simply specify the loss to be the (negative) ELBO and readily use a differentiable programming framework to perform optimisation. 
This is usually referred to as black-box variational inference \cite{blackboxVI}, automatic differentiation variational inference (ADVI) \cite{ADVI}, or (whenever we work with minibatches of the data) stochastic variational inference (SVI) \cite{SVI}.

\subsubsection*{Mean-field variational inference}

One simple choice of a variational family is the \textit{mean-field} variational family $Q = \{\normal(\params \mid  \boldsymbol{\mu}, \boldsymbol{\sigma}^T I) \mid \boldsymbol{\mu} \in \mathbb{R}^N,  \boldsymbol{\sigma} \in \mathbb{R}_+^N\}$, where $N$ is the size of the parameter space. In other words, the variational family consists of all independent multivariable Gaussian distributions. For example, consider the model from \autoref{fig:vi-example}:
$$x \sim \normal(0, 1) \qquad\qquad\qquad y \sim \normal(-2x^2, 1)$$
The mean-field variational family for this model is given by all distributions $q(x, y; \mu_{x,y}, \sigma_{x,y}) = \normal(x \mid \mu_x, \sigma_x)\normal(y \mid \mu_y, \sigma_y)$ for $\mu_{x,y} \in \mathbb{R}$ and $\sigma_{x,y} \in \mathbb{R}_+$.

\subsubsection*{Advantages and shortcomings}

Variational inference scales: it works for very large amounts of data, and can utilise modern frameworks and architectures to achieve fast inference. However, it trades accuracy to be able to do so. The variational family $Q$ is typically restricted to be potentially much simpler than the target distribution $p$, and we cannot, in general, determine how far off would even the optimal $q$ be from this target. \autoref{fig:vi-example} shows an example of a poor approximation resulting from variational inference.
The simple mean-field approximation is not able to capture the tails of the banana-shaped distribution, which results in a poor estimation of expectations under the inferred posterior: for example, the true expectation of $y$ is $-2$, and its standard deviation is $3$, but under the variational posterior, the expectation of $y$ is approximately $-0.4$ and its standard deviation is $1$.

\subsection{Inference on graphical models}

MCMC and VI both require a tractable density (or mass) function $p^*(\params)$, which is proportionate to the distribution of interest with respect to the parameters $\params$. 
However, it is not typical for these algorithms to make use of further information about the structure of the function $p^*(\params)$, such as its \textit{factorisation}. Information about the factorisation of the target distribution allows us to reason about conditional independencies between variables and develop algorithms that actively exploit the structure of a probabilistic model to make inference more efficient.

We express the factorisation of a distribution $p(\params) = \prod_{i}\phi_i(\params^{(i)})$ in terms of \textit{graphical models} (\citealt[Section 8.4]{Bishop}; \citealt[Part I]{Barber}; \citealt[Chapter 20]{Murphy}). Here, each $\phi_i$ is some function on a subset of the parameters $\params^{(i)} \subseteq \params$. A graphical model encodes this factorisation through a graph, where each parameter $\theta \in \params$ is a separate vertex, and dependencies between parameters appearing in the same subset $\params^{(i)}$ are expressed by connections between those vertices corresponding to $\params^{(i)}$. 
In particular, \textit{factor graphs} \cite{FactorGraphs} are bipartite graphs, with a \textit{variable vertex} for each $\theta \in \params$ and a \textit{factor vertex} for each factor $\phi_i(\params^{(i)})$ of $p(\params)$. An edge exists between a variable node $\theta$ and a factor $\phi_i(\params^{(i)})$ if and only if $\theta \in \params^{(i)}$.

For example, consider a Hidden Markov Model (HMM) with $N$ parameters $\mathbf{z} = (z_1, \dots, z_N)$, $N$ observed variables $\mathbf{y} = (y_1, \dots, y_N)$, and some given constant arrays $\boldsymbol{\alpha}$ and $\boldsymbol{\beta}$:
\begin{align*}
z_1 &\sim \mathrm{Bernoulli}(\alpha_{1}) & \\ 
z_n &\sim \mathrm{Bernoulli}(\alpha_{z_{n-1}}) &\text{ for n = 2,\dots, N} \\
y_n &\sim \mathrm{Bernoulli}(\beta_{z_n}) &\text{ for n = 1,\dots, N}
\end{align*}

Using $\bern(x \mid \pi)$ as a shorthand for the probability mass function of a $\mathrm{Bernoulli}(\pi)$ variable $x$, the factorisation of the joint distribution over $\mathbf{z}$ and $\mathbf{y}$ is:
\begin{equation} 
p(\mathbf{z}, \mathbf{y}) = \bern(z_1 \mid \alpha_1) \bern(y_1 \mid \beta_{z_1}) \prod_{n=2}^{N}  \bern(z_n \mid \alpha_{z_{n-1}}) \bern(y_n \mid \beta_{z_n})
\end{equation}

\begin{figure} 
	\centering
	\includegraphics[width=0.6\textwidth]{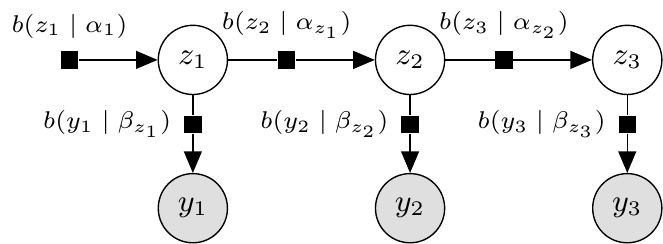}
	\caption{Factor graph corresponding to the factorisation of a simple HMM (Eq.~\ref{eq:hmm-factorisation}).}
	\label{fig:factor_graph}
\end{figure}

More concretely, let's say $N=3$. Then the joint is:
\begin{equation} \label{eq:hmm-factorisation}
	p(z_1,z_2,z_3,y_1,y_2,y_3) = \bern(z_1 \mid \alpha_1) \bern(z_2 \mid \alpha_{z_{1}}) \bern(z_3 \mid \alpha_{z_{2}}) \\ \bern(y_1 \mid \beta_{z_1})  \bern(y_2 \mid \beta_{z_2}) \bern(y_3 \mid \beta_{z_3})
\end{equation}
\autoref{fig:factor_graph} shows this expressed in terms of a factor graph: round vertices are variable nodes and square vertices are the factor nodes, $\bern(z_1 \mid \alpha_1), \bern(y_1 \mid \beta_{z_1})$ and so on. Shaded variable nodes ($y_1, y_2$ and $y_3$) denote observed variables.

Suppose we want to compute the posterior $p(\mathbf{z} \mid \mathbf{y}) = \frac{p(\mathbf{z}, \mathbf{y})}{\sum_{z} p(\mathbf{z}, \mathbf{y})}$. Calculating this in the general case requires $N$ nested sums over the parameters $\mathbf{z}$, meaning its complexity is exponential: $O(2^N)$
But knowing the factorisation of the joint $p(\mathbf{z}, \mathbf{y})$ allows us to rewrite the expression for computing the normalising constant $Z = \sum_{z} p(\mathbf{z}, \mathbf{y})$:
\begin{align*}
Z &= \sum_{z} p(\mathbf{z}, \mathbf{y}) \\
&= \sum_{z_1}\sum_{z_2}\sum_{z_3} \bern(z_1 \mid \alpha_1) \bern(z_2 \mid \alpha_{z_{1}}) \bern(z_3 \mid \alpha_{z_{2}}) \bern(y_1 \mid \beta_{z_1})  \bern(y_2 \mid \beta_{z_2}) \bern(y_3 \mid \beta_{z_3}) \\
&= \sum_{z_3} \bern(y_3 \mid \beta_{z_3})
\left[\sum_{z_2}  \bern(z_3 \mid \alpha_{z_{2}})  \bern(y_2 \mid \beta_{z_2}) 
\left[\sum_{z_1} \bern(z_1 \mid \alpha_1) \bern(z_2 \mid \alpha_{z_{1}}) \bern(y_1 \mid \beta_{z_1}) \right] \right]
\end{align*}

The inner-most sum over $z_1$ only depends on $z_2$, and thus we can compute it independently of $z_3$ as a function of $z_2$. This generalises beyond the particular value of $N$ we chose:
\begin{align*}
Z = \sum_{z_N} \bern(y_N \mid \beta_{z_N}) & \left[ \sum_{z_{N-1}}  \bern(z_N \mid \alpha_{z_{N-1}}) \bern(y_{N-1} \mid \beta_{z_{N-1}}) \dots \right.  \\
& \left. \left[\sum_{z_2}  \bern(z_3 \mid \alpha_{z_{2}})  \bern(y_2 \mid \beta_{z_2}) 
\left[\sum_{z_1} \bern(z_1 \mid \alpha_1) \bern(z_2 \mid \alpha_{z_{1}}) \bern(y_1 \mid \beta_{z_1}) \right] \right] \dots \right] 
\end{align*}

Starting from the inner-most expression $\left(\sum_{z_1} \bern(z_1 \mid \alpha_1) \bern(z_2 \mid \alpha_{z_{1}}) \bern(y_1 \mid \beta_{z_1})\right)$, we reduce each bracket to a function of a single variable $z_{n+1}$, while summing out another variable, $z_n$. In other words, it takes $O(2^2)$ time to compute each of $N-1$ intermediate expressions, thus the overall complexity is $O(N)$. A huge improvement over the $O(2^N)$ from before!

Inference algorithms based on graphical models can utilise the structure of a given model to automatically derive an efficient model-specific inference strategy, as we did by hand above. Some algorithms, such as the sum-product algorithm \cite[Section 8.4.4]{Bishop} are exact, but restricted to tree-structured graphical models in which computing exact marginals is tractable. Others, like variational message-passing \cite{VMP} and expectation propagation \cite{EP}, as used by Infer.NET \cite{InferNET}, are applicable to a large class of graphical models, but only provide an approximate solution. Here, we discuss in more detail one algorithm for exact inference that relates to the work described in \autoref{ch:slicstan2}: variable elimination.   

\subsubsection*{Variable elimination}

\textit{Variable elimination} (VE) \cite{zhang1994simple, koller2009probabilistic} is an exact inference algorithm efficient in models with sparse structure. It is applicable to general graphs, but it requires the model to be such that marginals are tractable. In particular, it applies to any model that contains only discrete parameters of finite support. 

The idea is to eliminate (marginalise out) variables one by one. To eliminate a variable $z$, we multiply all of the factors connected to $z$ to form a single expression, then sum over all possible values for $z$ to create a new factor, remove $z$ from the graph, and finally connect the new factor to all former neighbours of $z$. Here, ``neighbours'' refers to the variables which are connected to a factor that connects to $z$.

\begin{figure*}
	\centering 
	\begin{subfigure}{0.45\textwidth} 
		\vspace{9.5pt}
		\hspace{-7pt} 
		\includegraphics[width=\textwidth]{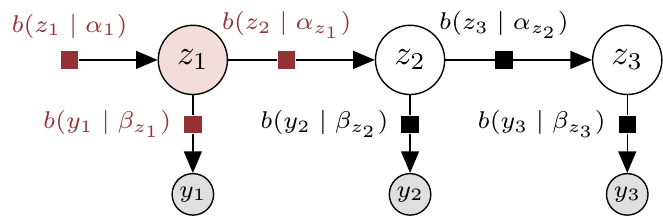}
		\caption{Remove $z_1$ and its neighbouring factors (red). Create a new factor $f_1$ by summing out $z_1$ from the product of these factors.}
		\label{fig:varelim_a}
	\end{subfigure}
	\hspace{0.08\textwidth} 
	\begin{subfigure}{0.45\textwidth} 		
		\includegraphics[width=\textwidth]{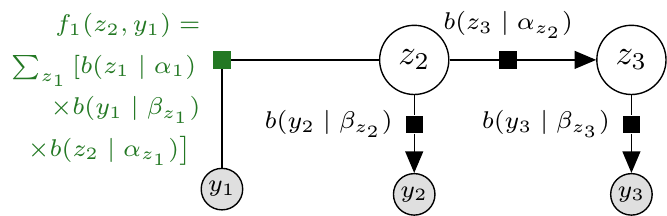}
		\caption{Connect $f_1$ (in green) to the former neighbours of $z_1$. The remaining factor graph defines the marginal $p(z_2, z_3, \mathbf{y})$.}
		\label{fig:varelim_b}
	\end{subfigure}
	
	\begin{subfigure}{0.45\textwidth}
		\hspace{-7pt} 
		\includegraphics[width=\textwidth]{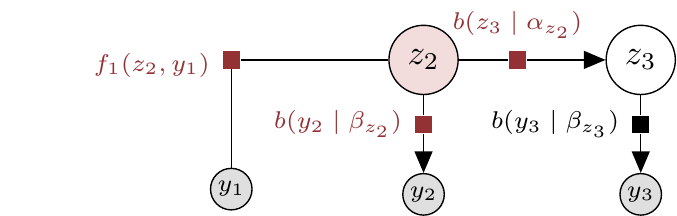}
		\caption{Remove $z_2$ and its neighbouring factors (red). Create a new factor $f_2$, by summing out $z_2$ from the product of these factors.}
		\label{fig:varelim_c}
	\end{subfigure}	
	\hspace{0.08\textwidth}
	\raggedright 
	\begin{subfigure}{0.45\textwidth} 
		\vspace{9pt}
		\includegraphics[width=\textwidth]{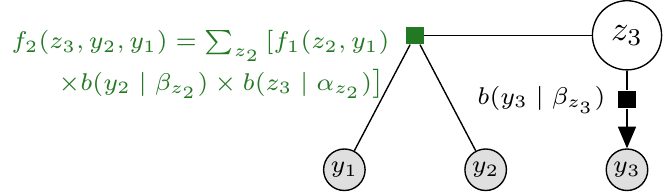}
		\caption{Connect $f_2$ (in green) to the former neighbours of $z_2$. The remaining factor graph defines the marginal $p(z_3, \mathbf{y})$.}
		\label{fig:varelim_d}
	\end{subfigure}
	\caption{Step by step example of variable elimination.}
	\label{fig:varelim}
\end{figure*}

\autoref{fig:varelim} shows the VE algorithm step-by-step applied to the HMM of length $3$ from before. We eliminate $z_1$ to get the marginal on $z_2$ and $z_3$ (\ref{fig:varelim_a} and \ref{fig:varelim_b}), then eliminate $z_2$ to get the marginal on $z_3$ (\ref{fig:varelim_c} and \ref{fig:varelim_d}). Finally, having obtained $p(z_3, \mathbf{y})$, we can sum out $z_3$ to obtain the normalising constant $Z$, which gives us the posterior $p(\mathbf{z} \mid \mathbf{y}) = \frac{p(\mathbf{z}, \mathbf{y})}{Z}$.

While the procedure shown in \autoref{fig:varelim} is optimal for this model, this would not have necessarily been the case if we had eliminated the variables in a different order (for example, starting from $z_2$ would be less efficient). VE provides an optimal exact marginalisation solution, but only with respect to a particular elimination ordering. Finding the optimal ordering is an NP-complete problem \cite{Yannakakis81, Arnborg87}, however many good heuristics exist (for example minimal degree ordering \cite{ordering-min-degree}).

\subsubsection*{Advantages and shortcomings}

In the cases when a model has a sparse structure, consisting of factors that allow for analytical marginalisation, variable elimination provides an efficient exact solution. However, in the general case, the complexity of VE is exponential in the number of nodes. 

More generally, algorithms that operate on graphical models include a large number of approximate inference techniques. These include loopy belief propagation \cite{BP}, variational message-passing \cite{VMP}, expectation propagation \cite{EP}, and other message-passing algorithms. While more general and able to scale, these methods suffer from the same problem we saw in \autoref{ssec:vi}: the approximated solution might be very far off the true distribution of interest.

\subsection{Other inference strategies}  

The list of algorithms presented so far is by no means exhaustive. The number of available Bayesian inference strategies, each coming with its own set of constraints, advantages and limitations, is too vast for it to be practical to include a full review in this dissertation. Here, I briefly discuss a few other strategies that are less relevant to the contributions of this work, but nevertheless provide a useful background when contemplating the challenges of probabilistic programming.  

\vspace{-4pt}
\subsubsection*{Sequential Monte Carlo}

Sequential Monte Carlo (SMC) \cite[Chapter~6]{SMC, SMC2, RainforthDiss}, also referred to as \textit{particle filtering}, is a Monte Carlo method that is particularly suitable for dynamic models that involve stochastic control flow and recursion. 
Like MCMC, SMC is a Monte Carlo algorithm that approximates a distribution through a collection of samples. While MCMC obtains such a collection by repeatedly mutating a sample from the distribution, the core idea behind SMC is to independently sample from some proposal distribution, and compute a weight associated with each such sample, based on how likely it is that the sample is also from the target distribution. 
Expectations under the target can then be computed as a weighted average of the obtained samples.

What makes SMC different to simple \textit{importance sampling} (\citealt[Section~27.6]{Barber}; \citealt[Section~11.1.4]{Bishop}), is that this sample-and-weight process is divided into stages. The algorithm sequentially samples each random variable given the already sampled variables and updates the weight. This has the advantage that at every intermediate step we can augment the population of samples 
by resampling them based on the current weights. This ensures that more samples are propagated in areas where we expect the probability mass of the target to be higher, thus making better use of available resources. 

SMC can also be used to construct a proposal distribution for MCMC methods. This combination between SMC and MCMC inference is referred to as \textit{Particle Markov chain Monte Carlo} \cite{PMCMC}.

SMC has been a popular algorithm in the field of probabilistic programming, due to its generality (see \autoref{sec:bg-ppls}). The structure made available by the presence of a probabilistic program is particularly useful for automatically devising a program-specific SMC inference algorithm, which works in the presence of stochastic control flow and is highly parallelisable. 
However, similar to other inference algorithms, SMC also comes with its disadvantages. In particular, the algorithm is sensitive to the choice of proposal distribution. A proposal distribution that is close to the true target would result in efficient inference, while a poorly-suited proposal would result in low effective sample size, especially in high dimensions (due to the curse of dimensionality, \autoref{ssec:curse}). This has resulted in research studying various techniques for improving the proposal, including  \textit{inference compilation}: learning a suitable proposal through variational inference \cite{InferenceCompilation}. SMC typically requires a number amount of samples, usually much larger than gradient-based methods like HMC, but it can be better suited for multi-modal distributions and is more general. 

For an excellent review of sequential Monte Carlo, refer to \citet[Chapter~6]{RainforthDiss}.

\subsubsection*{Likelihood-free inference methods}

Most of the methods discussed so far are \textit{likelihood-based}. That is, they explicitly evaluate the likelihood $p(\data \mid \params)$ to perform inference. However, in some cases, the likelihood is unavailable or intractable. \textit{Likelihood-free} inference methods
can perform Bayesian inference based solely on simulation from the generative model. 
Likelihood-based and likelihood-free are umbrella terms that include many inference strategies, and thus lack canonical references, but have been recently surveyed, for example, by \citet{PapamakariosDiss} and \citet{CranmerLFI}.

We already saw one likelihood-free inference method: rejection sampling (\autoref{ssec:curse}). Rejection sampling is perhaps the simplest likelihood-free inference algorithm, where we simulate parameters $\hat{\params}$ and data $\hat{\data}$ from the model and throw away (or \textit{reject}) any simulated data $\hat{\data}$ that does not \textit{exactly} match the observed data $\data$. 
This is likelihood-free, as we do not evaluate the likelihood function at any point. 
Approximate Bayesian computation (ABC) \cite{ABC, ABC2} methods are based on a similar idea, but accept simulated data that is sufficiently close, some small distance $\epsilon$, from the true data. That is, a sample $\hat{\params}$ is accepted whenever $|\hat{\data} - \data| < \epsilon$.
Many likelihood-based Monte Carlo methods can be augmented to a likelihood-free alternative, by approximating the likelihood this way, which gives rise to algorithms such as MCMC-ABC \cite{MCMC-ABC} and SMC-ABC \cite{SMC-ABC}. These algorithms are more general, as they do not require a tractable likelihood, but are less efficient than their likelihood-based counterparts.

Recent work in the area also includes likelihood-free inference through \textit{neural density estimation} \cite{Durkan2018, SNL, PapamakariosFlows2019,NeuralLFI}. 
Density estimation is the problem of estimating the value of the probability density function of some random variable at an arbitrary point, given only a set of independently generated samples of that variable; 
neural density estimation tackles this by using neural networks.  
This relates closely to the problem of likelihood-free inference, where we can generate such a set of samples, but do not know the density function of the data. 
Sequential neural methods for likelihood-free inference \cite{Durkan2018, NeuralLFI} perform inference by either 
approximating the likelihood function using neural density estimation and using it in combination with likelihood-based methods, or by directly estimating the posterior density. 

For an excellent review of likelihood-free inference, refer to \citet[Part II]{PapamakariosDiss}.

\section{Probabilistic programming languages} \label{sec:bg-ppls}

Probabilistic programming \cite{Gordon14, IntroductionPP2018} aims to democratise the difficult process of Bayesian inference by decoupling the process of writing a model from the actual inference algorithm. However, we saw in \autoref{sec:bg-inference} that inference is not an easy task. What algorithm is best suited for inference in a given model depends hugely on the properties of the model itself. Does the model have a fixed number of random variables? Does the model contain only discrete variables, only continuous variables, or both? Is the model differentiable? 
There does not exist a single inference algorithm that provides an efficient solution for all possible models. 

Probabilistic programming languages (PPLs) inherit this difficulty. Usually, languages trade off generality for efficiency of inference or automation. Some PPLs would restrict the space of supported models in order to use an efficient algorithm such as HMC. Others support a more general class of models, but are then forced to use a general inference algorithm, which can be less efficient than less general alternatives. Some PPLs have adopted the concept of programmable inference, which allows for the best of both worlds, but requires significantly more effort from the user, who now needs to choose and adapt the inference algorithm themselves. 

This leads to the development of many different PPLs, each with its own set of advantages and constraints. 
This section aims to provide a brief overview of the topic of probabilistic programming languages. 
The area has significantly expanded in recent years, and thus covering all existing PPLs in this dissertation will not be practical, but I focus on a few of the more popular languages, as a basis for the discussion on program analysis. 

\subsection{Definition and classification of PPLs}

Recall the simple probabilistic program from \autoref{sec:bg-simplePP}:
\begin{lstlisting}
	c1 ~ bernoulli(0.5)
	c2 ~ bernoulli(0.5)
	bothHeads = c1 and c2
	observe(not b)
\end{lstlisting}

There are two things that make this program different to a conventional program: the probabilistic assignment (or sample), $\sim$, and the observe statement, \kw{observe}. This follows perhaps the most established and most often used definition of probabilistic programs, which is given by \citet{Gordon14}:
\begin{quote}
	Probabilistic programs are ``usual'' programs (...) with two added constructs: (1) the ability to draw values at random from distributions, and (2) the ability to condition values of variables in a program via observe statements.
\end{quote}

While all PPLs fit this definition, the details can vary significantly. Both sampling and observation are understood and implemented in different ways across languages, yielding vastly different syntaxes and workflows. How do we classify probabilistic programming languages so that we can easily compare and contrast them?
In the following sections, I give one way to distinguish and talk about PPLs in a structured way. However, it is important to note that not all PPLs will fall in one of the described categories, and the described classification should not be seen as a formal taxonomy.

A fundamental way in which PPLs differ is their treatment of probabilistic assignment ($\sim$). There are (at least) two ways in which we can interpret a statement such as $y \sim \mathrm{dist}(x)$: (1) as stating that the underlying joint probability density contains the factor $\mathrm{dist}(y \mid x)$, or (2) as a random draw from the distribution $\mathrm{dist}(x)$ the result of which is bound to $y$. 
Importantly, these two are \textit{not} equivalent treatments. 
For example, consider the simple program $x \sim \mathrm{normal}(0, 1); x \sim \mathrm{normal}(0, 1)$. The unnormalised joint density under (1) is over a single variable, $x$: $p^*(x) = \normal(x \mid 0, 1)\normal(x \mid 0, 1)$. But in the case of (2), we sample, one by one, \textit{two} separate variables $x$, say we denote them $x_1$ and $x_2$. The corresponding unnormalised density of the program is $p^*(x_1, x_2) = \normal(x_1 \mid 0, 1)\normal(x_2 \mid 0, 1)$. As these variables are bound to the same variable name $x$, the first result is overwritten. We are left with the result of only the second sample statement, thus, we can also see the program as corresponding to the density $p(x_2) = \int p(x_1, x_2) \dif x = \normal(x_2 \mid 0, 1)$.

Programs written in languages that adopt the former, \textit{density-based} interpretation, explicitly define an (unnormalised) joint density function. The (unnormalised) likelihood of the model is usually available, it can be evaluated and used to perform likelihood-based inference. However, sampling data from the model is not always possible. Such models are called \textit{explicit} or \textit{prescribed} \cite{ImplicitVsExplicit}. 

PPLs that adopt the latter, \textit{sample-based} interpretation yield programs that are effectively simulators. If ran ignoring all observe statements, such programs have the effect of simulating data from the model. Models that are defined in terms of a sampling procedure are called \textit{implicit} \cite{ImplicitVsExplicit}, as we cannot, in general, evaluate the likelihood function of these models: it is implicitly defined.

While there are many categories in which we can divide PPLs, including based on programming style (declarative, imperative, or functional) or type of inference used (Monte Carlo, symbolic, variational, etc.), here I propose a way of classifying them based on their interpretation of $\sim$. 
This gives rise to three categories: \textit{density-based} (or \textit{explicit}) PPLs (\autoref{ssec:bg-density-ppls}) that yield explicit models, \textit{sample-based} (or \textit{implicit}) PPLs that yield implicit models (\autoref{ssec:bg-trace-ppls}), and \textit{effect-handling-based} PPLs (\autoref{ssec:bg-eff-ppls}), which can interpret $\sim$ either explicitly or implicitly based on context.
This classification is useful for several reasons: 
\begin{enumerate}
\item It relates PPLs and the models expressible in those PPLs to the established and useful distinction between explicit and implicit probabilistic models \cite{ImplicitVsExplicit}. It is important to note that these two model categories are not mutually exclusive. It is possible to simulate data from some density-based programs, and it is also possible to evaluate the likelihood for some sampling-based programs. 
\item It directly points to a suitable way of formalising the semantics of the language.  Typically the semantics of a PPL can be formalised in several equivalent ways, however, the distinction between explicit and implicit PPLs is helpful when the aim is to develop semantics that is closely tied to implementation. 
Explicit PPLs can be given straightforward \textit{density-based semantics}, which is the focus of \autoref{ch:slicstan}. Implicit PPLs are better formalised through \textit{trace-based} (or \textit{sampling-based}) \textit{semantics} (for example as described by \citet{Hur15}, \citet{StatonSemantics16}, \citet{QBS}).
\item It hints at the class of algorithms suitable for each language. Being able to evaluate an (unnormalised) density, defined on a fixed parameter space, is essential for gradient-based algorithms such as HMC or VI, making these algorithms easily applicable to explicit models. On the other hand, an implicit model will sometimes require a more general inference algorithm, such as SMC, or in some cases, even a likelihood-free algorithm (\citet[Section~3]{CranmerLFI} give details on when likelihood-based methods are applicable to a probabilistic program).
\end{enumerate}

The following sections present one or two example probabilistic programming languages for each of the three PPL types. In particular, I give an overview of Stan \cite{StanJSS} and Edward2 \cite{Edward2} (among other PPLs), which are PPLs central to \autoref{part:static} and \autoref{part:dynamic} of this dissertation respectively.

\subsection{Density-based (or \textit{explicit}) PPLs} \label{ssec:bg-density-ppls}

Density-based PPLs, which I am also going to refer to as explicit PPLs, explicitly define an unnormalised joint probability density. 
Examples of density-based languages include Stan \cite{StanJSS}, BUGS \cite{BUGS}, and JAGS \cite{JAGS}, as well as languages that operate on graphical models, including Infer.NET \cite{InferNET}, Fun \cite{Fun}, and Tabular \cite{Tabular}. 
Turing \cite{Turing}, a PPL that allows for compositional inference with different inference techniques being used for different sub-parts of the model, is also density-based.

In these languages, statements such as $y \sim \mathrm{dist}(x)$ are interpreted as information that the joint density given by the model contains the factor $\mathrm{dist}(y \mid x)$. The parameter space is fixed: it consists of some parameters of interest $\params$, such that the dimension of $\params$ does not change during inference. A model in a density-based PPL usually describes a \textit{tractable} unnormalised density $p^*(\params, \data)$; in other words, the model can be used to evaluate $p^*(\params, \data)$ for different parameter values $\params$. Languages that support \textit{deterministic observations} (that is \lstinline{observe($f(\data) = \hat{\data}$)} for some non-identity deterministic function $f$, as opposed to \lstinline{observe($\data = \hat{\data}$)}), such as Infer.NET, are an exception to this tractable unnormalised density rule. 

This constrained interpretation of $\sim$ leads to several advantages of density-based PPLs. 
As the parameter space is fixed, 
and the (unnormalised) joint density is usually available in closed form and can be evaluated, 
explicit PPLs are, in general, able to use more efficient algorithms compared to sample-based PPLs. In particular, we can apply gradient-based algorithms such as HMC or VI. 
It is also intuitive to write both directed and undirected models, as probabilistic statements are understood simply as adding a multiplicative factor to the target density. 
However, it is not always possible to efficiently simulate data from the model. Explicit models are also more constrained in terms of recursion and conditionals, as the number of random variables they need is fixed. 

Below, I briefly discuss two density-based PPLs: Stan and Infer.NET. 

\subsubsection*{Stan}

Stan \cite{StanJSS} is a popular probabilistic programming language. It has been adopted and used in numerous fields, including 
cosmology \cite{StanCosmology},
microbiology \cite{StanBiostats},
epidemiology  \cite{StanCOVID1, StanCOVID2},
sociology \cite{StanSociology},
psycho-linguistics \cite{StanPsychoLinguistics},
and business forecasting \cite{StanForscasting}.
Stan's syntax is inspired by and similar to that of BUGS \cite{BUGS}, and is close to the model specification conventions used in publications in the statistics community.
Stan is imperative and compiles to C++ code. It consists of program blocks, which contain sequences of variable declarations and statements.
For example,
consider the \textit{Eight schools model}, defined as:
\vspace{-2pt}
\begin{align*}
	\mu &\sim \mathcal{N}(0, 5) \qquad \tau \sim \mathrm{HalfCauchy}(0, 5) \\
	\theta_n &\sim \mathcal{N}(\mu, \tau) \qquad
	y_n \sim \mathcal{N}(\theta_n, \sigma_n) \qquad \text{for } n = 1,\dots, 8
\end{align*}
\vspace{-2pt}
where $N, y_n, \sigma_n$ for $n \in \{1, ..., N\} $ are given as data.

\autoref{prog:stan8schools} gives the Stan program for Eight schools. It declares the observed data $N, \mathbf{y}$, and $\boldsymbol{\sigma}$ in the \kw{data} block and, separately, the parameters $\mu, \tau,$ and $\boldsymbol{\theta}$ in the \kw{parameters} block. As $\tau$ comes from a half-Cauchy distribution,
\footnote{The half-Cauchy distribution has probability density function proportionate to the probability density function of the Cauchy distribution, but the support of half-Cauchy is constrained to $\mathbb{R}^+$.}
it takes only positive values: the way to write this in Stan is through a \textit{constrained type}, in this case,  \kw{real<lower=0>}. Finally, the \kw{model} block defines the unnormalised joint density of the model. Formally, that is the model defines $p(\mu, \tau, \boldsymbol{\theta}, \boldsymbol{\sigma}, \mathbf{y}) \propto \normal(\mu \mid 0, 5) \mathrm{HalfCauchy}(0, 5) \prod_{n=1}^{N}\normal(\theta_n \mid \mu, \tau) \normal(y_n \mid \theta_n \sigma_n)$. The inference task is to find $p(\mu, \tau, \boldsymbol{\theta} \mid \boldsymbol{\sigma}, \mathbf{y})$.

In general, a Stan program can contain up to seven blocks, with all blocks discussed in detail in \autoref{ch:slicstan}. Most notably, the \kw{data} block declares the data $\data$ and the \kw{parameters} block declares the parameters $\params$. Executing these blocks in order by treating the code as a conventional imperative code results in evaluating the log density $\log p(\data, \params)$ for the observed data $\data$ and at some value of the parameters $\params$.

\begin{wrapfigure}{L}{0.46\textwidth}
\vspace{-22pt}
\begin{lstlisting}[caption={Eight schools in Stan}\label{prog:stan8schools}, escapechar=@]
 data {
	 int N;
	 real y[N];
	 real sigma[N];
 }
 parameters {
	 real mu;
	 real<lower=0> tau;
	 real theta[N];
 }
 model {
	 mu ~ normal(0, 5);
	 tau ~ cauchy(0, 5);
	 for (n in 1:N) {
  	theta[n] ~ normal(mu, tau);
  	y[n] ~ normal(theta[n], sigma[n]);
	 }
 }@\vspace{-5pt}@
$\fooline$
\end{lstlisting}
\vspace{-42pt}
\end{wrapfigure}

\vspace{-8pt}
\subparagraph{Inference.}
Stan uses HMC, and more specifically, an enhanced version of the No-U-Turn Sampler (NUTS) \cite{NUTS,HMCConceptual}, which is an adaptive path lengths extension to HMC. 
As described above, a Stan program essentially defines the body of a function that takes as an argument the parameters $\params$ and evaluates the joint log density at that point $\params$. As $p(\params \mid \data) \propto p(\params, \data)$, we can use this function for MCMC inference.  
Stan performs inference by running the code imperatively for different parameter values while using automatic differentiation to compute gradients with respect to the parameters.

\vspace{-8pt}
\subparagraph{Strengths.}

Stan is flexible, in that it gives the user full control over the definition of the target joint density.
Compilation of Stan code goes through several stages of optimisation and static transformations, which increases runtime performance and stability of inference. 
Inference is fully automatic, including HMC hyperparameter tuning and automatic differentiation, as well as very efficient. 
Stan also automatically performs numerous diagnostic checks upon inference and issues warnings and suggestions for improvement.

\vspace{-6pt}
\subparagraph{Constraints.}

Being density-based, Stan requires a fixed number of parameters during inference. It is also not generally possible to simulate data from the model.
In addition, 
the block syntax makes it difficult to compose different Stan programs, or to have flexible user-defined functions (which I address in \autoref{ch:slicstan}).
HMC requires the joint density to be piece-wise differentiable. This means having discrete \textit{data} in Stan is possible, but it is not possible to explicitly encode discrete \textit{parameters} (which I address in \autoref{ch:slicstan2}). 
While the Stan compiler performs some static optimisations, statistical optimisations, such as reparameterisation and marginalisation, are mostly the responsibility of the user.

\vspace{-2pt}
\subsubsection*{Infer.NET}
\vspace{-1pt}
Infer.NET \cite{InferNET} is a probabilistic programming framework, which can be used within the .NET ecosystem, and it is well-known for its use in skill rating systems \cite{TrueSkill}, recommendation systems \cite{InferNETRecSys}, and other applications that require real-time responsiveness. 
Infer.NET constructs an explicit factor graph corresponding to the specified program. Every program variable is a variable node in that graph, and every statement, probabilistic or deterministic, is a factor in that graph. 
\autoref{prog:infernet8schools} shows the Eight schools example written in Infer.NET.

\begin{lstlisting}[caption={Eight schools in Infer.NET}\label{prog:infernet8schools}, escapechar=@]
 var N = Variable.New<int>();
 var NRange = new Range(N);
 var sigma = Variable.Array<double>(NRange);
 var mu = Variable.GaussianFromMeanAndPrecision(0, 5);
 var tau_sq = HalfCauchySquared(Variable.Constant(12.5));
 var theta = Variable.Array<double>(NRange);
 theta[NRange] = Variable.GaussianFromMeanAndPrecision(mu, tau_sq).ForEach(NRange);
 var y = Variable.Array<double>(NRange);
 y[NRange] = Variable.GaussianFromMeanAndPrecision(theta[NRange], sigma[NRange]);@\vspace{-9pt}@
$\fooline$
\end{lstlisting}
\vspace{-20pt}

\subparagraph{Inference.} 
Infer.NET programs compile to efficient message-passing code. The language supports both \textit{expectation propagation} \cite{EP} and \textit{variational message-passing} \cite{VMP}, both of which provide an approximate solution. In models whose factor graph is a tree of discrete variables, Infer.NET's message-passing algorithm reduces to \textit{belief propagation} \cite{BP}, which is equivalent to variable elimination (\autoref{sec:bg-inference}) and gives an exact solution. 
For some models, Infer.NET is able to exploit the conditional independence structure available through the explicit factor graph, and use \textit{blocked Gibbs sampling} \cite{Gibbs}.

\subparagraph{Strengths:}

One of the biggest advantages of Infer.NET is its speed of inference: the language has been employed in real-life systems that work with large amounts of data and require very fast inference results.
In addition, inference is also fully black-box.

\subparagraph{Constraints:}

In addition to the constraints of a density-based PPL, Infer.NET also supports only certain operators. For example, division and multiplication between variables is not always possible, and the compiler can sometimes throw an error requesting the model to be reformulated.
Unlike other PPLs, Infer.NET also does not provide results about the joint posterior distribution of parameters, but can only infer the marginal distributions.

\subsection{Sample-based (or \textit{implicit}) PPLs} \label{ssec:bg-trace-ppls}

Sample-based PPLs, or also implicit PPLs, define the unnormalised joint density only implicitly, by encoding a way to sample from that density. Examples of such languages include Anglican \cite{Anglican, AnglicanTolpin}, Church \cite{Church}, Gen \cite{Gen}, WebPPL \cite{WebPPL}, and PyProb \cite{PyProb}. 
In addition, likelihood-free inference PPLs, such as ELFI \cite{ELFI}, SBI \cite{SBI}, and Omega \cite{Omega} are also sample-based. That is because they define the joint density only implicitly through simulation, and support models where the likelihood is not available by definition. 

In these languages, statements such as $y \sim \mathrm{dist}(x)$ are understood as drawing a sample from the distribution $\mathrm{dist}(x)$ at random and binding it to $y$. As a consequence, models in sample-based PPLs are essentially \textit{simulators}: they describe the process of generating data. When ran forward, these models can usually give us samples from the joint density defined by the model. However, evaluating this joint density on a fixed-size subset of the parameters is not always possible.  For example, suppose we are interested in parameters $\params$ that are of fixed support and that the model defines $\params$ and some set of latent variables $\mathbf{z}$ that have variable support. Then the unnormalised density $p^*(\params, \data) \propto \int p(\params, \mathbf{z}, \data) \dif \mathbf{z}$ is not always tractable.
It is in this sense that we say that the density is \textit{implicit}: $p^*(\params, \data)$ is not always tractable, even though an explicit density on \textit{traces} $p(\params, \mathbf{z}, \data)$ may still exist.  

Sample-based PPLs often have cleaner, more interpretable syntax compared to density-based PPLs, but need to adopt less efficient, more general-purpose inference algorithms. Sample-based PPLs support a larger class of models, including models with complicated control-flow, recursion, and unbounded number of random variables. While undirected models are also often supported, they are, in general, less intuitive to write than in density-based PPLs, due to lack of generative interpretation for undirected factors.

Below, I once again use the Eight schools example to give brief overview of two sample-based PPLs: Anglican and Gen.

\subsubsection*{Anglican} 

\begin{wrapfigure}{L}{0.55\textwidth}
\vspace{-22pt}
\begin{lstlisting}[caption={Eight schools in Anglican}\label{prog:anglican8schools},style=Anglican]
(defquery schools [N, y, sigma]
	(let [mu (sample (normal 0 5))
	      tau (sample (gamma 1 1))]
	 (loop [n 0]
	   (if (< n N)
	     (let [y_n (nth y n)
	           sigma_n (nth sigma n)
	           theta_n (sample (normal mu tau))]
	       (observe (normal theta_n sigma_n) y_n)
	       (recur (+ n 1)))))
	[mu, tau]))
$\fooline$
\end{lstlisting}
\vspace{-32pt}
\end{wrapfigure}

Anglican \cite{Anglican, AnglicanTolpin} is a general-purpose PPL integrated in Clojure. 
It supports a wide set of probabilistic programs, including programs with discrete and continuous parameters, unbounded recursion and control flow. 
The language has been extensively formalised and has strong theoretical foundations \cite{StatonSemantics16}.

\autoref{prog:anglican8schools} shows Eight schools written in Anglican.
\footnote{This Anglican version of Eight schools has been given slightly different priors compared to the previously introduced model, due to availability of built-in distributions of the language.} 
The model is a \lstinline[style=Anglican]{defquery}: a function that takes as input the observed data and returns the parameters of interest. Internally, this query is transformed to continuation passing style (CPS) (similarly to WebPPL \cite{WebPPL}), where continuations are explicitly maintained at two \textit{checkpoints}: \lstinline{sample} and \lstinline{observe}. In brief, an Anglican program is a standard Clojure program, with the exception of \lstinline{sample} and \lstinline{observe}, which are executed differently, as defined by the inference algorithm. Each inference algorithm defines its own implementation for the two checkpoints and is able to alter the behaviour of the code to produce samples from the posterior. \citet[Chapter~7]{RainforthDiss} discusses inference algorithm implementation in Anglican in detail, while below I discuses the inference strategies readily available as part of the language.

\subparagraph{Inference:}

Anglican supports a range of inference techniques. 
To allow for inference in any program expressible in the language, Anglican  was the first PPL to adopt particle-based inference \cite{RainforthDiss}.
In addition to simple importance sampling, standard SMC \cite{SMC,SMC2} and PMCMC \cite{PMCMC}, the language also introduces several new and PPL-focused particle-based techniques, such as interacting particle MCMC \cite{iPMCMC} and particle Gibbs with ancestor sampling \cite{PGAS}.

\subparagraph{Strengths:}
Anglian provides clean functional syntax for general probabilistic programming. The semantics of the language is well-studied and give it a strong theoretical background. Anglican supports higher-order programs with varying support, recursion and stochastic control-flow. 
It supports a large number of fully automated inference algorithms, and is perhaps the most sophisticated PPL in terms of particle-based inference. 
Composing models in Anglican is straightforward.

\subparagraph{Constraints:}
Anglican inherits some of the limitations of the inference algorithms it adopts. 
In order to be well-suited for SMC inference, an Anglican program needs to interleave \lstinline{sample} and \lstinline{observe} statements as much as possible. This is in direct contrast to other languages, where in general it does not matter at what point observations are made. 
To ensure high-quality inference, particle-based methods require a large number of samples, and thus are typically more computationally-demanding than alternatives.

\subsubsection*{Gen} 

Gen \cite{Gen} is a Julia-based PPL centred around the idea of programmable inference \cite{ProgrammableInference}. 
It consists of components that allow the user to build their own inference algorithms, for example by building custom MCMC kernels, or combining several inference strategies. Like Anglican, Gen is sample-based and supports programs with complex control flow and recursion.

\begin{wrapfigure}{L}{0.57\textwidth}
\vspace{-22pt}
\begin{lstlisting}[caption={Eight schools in Gen}\label{prog:gen8schools},style=Gen]
@gen function schools(N::Int, $\sigma$::Vector{Float64})
	$\mu$ = @trace(normal(0, 5), :mu)
	$\tau$ = @trace(gamma(1, 1), :tau)
	for (n, $\sigma$_n) in enumerate($\sigma$)
		$\theta$_n = @trace(normal($\mu$, $\tau$), (:theta, n))
		y_n = @trace(normal($\theta$_n, $\sigma$_n), (:y, n))
	end
	return nothing
end;
$\fooline$
\end{lstlisting}
\vspace{-32pt}
\end{wrapfigure}

\autoref{prog:gen8schools} gives the Eight school model written in Gen. The model is a Julia function defined using the annotation \lstinline[style=Gen]{@gen}. The function takes as arguments the \textit{unmodelled data} (data we do not assign probability to) $N$ and $\boldsymbol{\sigma}$, and defines all modelled variables using a \lstinline[style=Gen]{@trace} expression and giving them unique addresses. A \lstinline[style=Gen]{@gen} function can be used to generate \textit{traces}: instances of a trace abstract data type, which contains random choices made in the program and provides various operators such as evaluating the log probability of the trace, or computing the gradient of the log joint density with respect to different parameters. 

\subparagraph{Inference:}

While Gen supports various built-in inference algorithms, such as variational inference, HMC and SMC, the biggest advantage of the language is the flexibility it provides in combining inference operators to build custom inference algorithms. Gen's API is particularly well-suited for exploring different SMC strategies and for defining custom MCMC kernels.  

\subparagraph{Strengths:}
 
Like other sample-based PPLs, Gen provides a very flexible modelling language that allows for an unbounded number of random variables. 
Composing probabilistic models in Gen is straightforward, and in addition, the language also provides \textit{generative function combinators} that act as higher-order functions, such as \kw{map}, and can compactly combine Gen models. 
Perhaps the biggest strength of Gen is that it provides components for building custom inference algorithms, while still automating lower-level details. 

\subparagraph{Constraints:}

While its biggest strength, programmable inference in Gen can be challenging for beginners. Using inference algorithms automatically is less straightforward than with other languages (for example, there is no fully automated version of SMC).
In addition, Gen's building blocks are constructed with specific focus on sampling-based and variational inference, and there is no support for algorithms such as variational message-passing, loopy belief propagation, and expectation propagation.
Gen supports transformations on traces, but it is not clear how to apply transformations that need program context, such as, for example, the one described in \autoref{ch:autoreparam}.

\subsection{Effect-handling-based PPLs} \label{ssec:bg-eff-ppls} 

Both density-based and sample-based PPLs assign a particular (and differing) meaning to the statement $y \sim \mathrm{dist}(x)$. In contrast, in effect-handling-based PPLs, this meaning is not fixed and can change depending on context. Such languages include Pyro \cite{Pyro}, NumPyro \cite{NumPyro}, Edward2 \cite{Edward2}, and Eff-Bayes \cite{OliverMScDiss}, and they use \textit{algebraic effect handlers} to specify the behaviour of a probabilistic program.

Algebraic effects and handlers \cite{Effects, Effects2} have emerged as a convenient, modular abstraction for controlling computational effects. Effect-handing-based PPLs tread the probabilistic assignment operation $\sim$ as an \textit{effectful operation}, the behaviours of which is specified by a separate \textit{effect handler}. Such an effect handler can specify behaviour matching that of a density-based PPL, or that of sampling-based PPL, but it can also give a completely different meaning to the $\sim$ operator. 

Defining and working with handlers is part of the PPL itself, and they can be nested to produce complex composable transformation \cite{ProbProg18}.
In contrast, other languages that treat $\sim$ as a special \textit{checkpoint} where program behaviour can be altered, such as
Anglican and WebPPL, can implement different algorithm-specific behaviour for sampling, but this is seen as part of the inference algorithm implementation rather than as part of the modelling language. 

I give more background on effect-handling-based PPLs, including examples of defining and composing handlers, in \autoref{sec:autoreparam-eff-ppls}. Here, I give a brief overview of one such PPL: Edward2.

\subsubsection*{Edward2}

Edward2 \cite{Edward2} is a \textit{deep} probabilistic programming language embedded into Python and built on top of TensorFlow \cite{Tensorflow}. 
It has found several applications in fields that require processing large amounts of data, such as
intelligent transportation systems \cite{Edward2Transport},
recommendation systems \cite{Edward2RecSim}, and
biological sequence models \cite{Edward2Biology}.
Edward2 uses effect handlers (referred to in the language as \textit{interceptors} or \textit{tracers}) to change the meaning of probabilistic assignment at runtime. Handlers can be defined by the programmer, and can also be nested to produce complex model transformations. 

\begin{lstlisting}[caption={Eight schools in Edward2}\label{prog:edward8schools},style=Edward]
	def schools(N, sigma):
		mu = ed.Normal(loc=0., scale=5., name="mu")
		tau = ed.HalfCauchy(loc=0., scale=5., name="tau")
		theta = ed.Normal(loc=mu * tf.ones(N), scale=tau * tf.ones(N), name="theta")
		y = ed.Normal(loc=theta, scale=sigma, name="y")
		return y
		
	with ed.interception(log_prob):	
		schools(N=8, sigma=data["sigma"])
$\fooline$
\end{lstlisting}

\autoref{prog:edward8schools} shows the Eight schools model in Edward2.
A model is simply a generative function, that creates some random variables. When run forward, it samples random variables using random number generation, meaning we can see Edward2 as a sampling-based PPL. However, when running the model in the context of a \lstinline{log_prob} handler (last two lines in \autoref{prog:edward8schools}), probabilistic assignment statements are instead treated as adding a term to the model's log joint density (similarly to Stan). Thus, in this case, we can think of Edward2 as a density-based PPL. 

\subparagraph{Inference:}

Edward2 focuses on gradient-based inference methods, and has automatic support for various algorithms such as variational inference and HMC. 
However, the flexibility provided by adopting effect handlers in the language makes it possible for the user to perform transformations on their model, compose inference strategies, or derive their own. I describe one such way in which effect handlers can be utilised to improve inference in \autoref{ch:autoreparam}.

\subparagraph{Strengths:}

Like other deep PPLs, Edward2's inference is particularly well-suited for many dimensional problems and large amounts of data. Composing models in the language is straightforward, and effect handlers provide a flexible way to perform programmable inference. In addition, handlers can be used as a tool for lightweight composable program transformation, which allows for model reparameterisation, marginalisation and automatic generation of variational families for VI \cite{ProbProg18}.

\subparagraph{Constraints:}

Edward2 requires some verbose and repeated syntax (for example, when naming variables). 
While providing flexibility, effect handlers can be challenging for inexperienced users and the automated inference support that Edward2 provides is not always as straightforward as with some alternatives. 
Caution is required when working with certain models; for example, models with discrete parameters need to be transformed to allow for gradient-based inference. 
Edward2's handlers are restricted to handlers where the program continuation is \textit{implicit} (see \autoref{ch:autoreparam} for details) and thus only a limited number of model transformations are possible.

\subsection{Other PPLs}

Classifying any complex subject into clean and well-defined categories is always hard, and this is also the case with the PPL classification suggested in this dissertation. There are languages that do not neatly fit in either of the categories we discussed. 

One major group of such languages is symbolic PPLs, where the inference result is an exact symbolic expression. 
For example, Hakaru \cite{Hakaru} and PSI \cite{PSI} transform probabilistic programs into symbolic representations, simplify those, and deliver an exact expression for the posterior, which may or may not be tractable. 
Dice \cite{Dice} compiles programs with discrete random variables into boolean decision diagrams. SPPL \cite{SPPL} is applicable to both discrete and continuous distributions and compiles programs to sum-product expressions.
The reason why these languages are difficult to categorise is that it is unclear whether we can consider the likelihood available or not. On the one hand, an expression for the likelihood might exist, and the language may be able to manipulate it to produce the final result. But on the other hand, the evaluation of the likelihood may not tractable, which is often the case when deterministic observations or unbounded number of random variables is allowed.

%% file: chapters/part1-intro.tex
\section*{Introduction to Part I}

\textit{Static analysis} refers to the class of techniques that can analyse programs \textit{without} executing them --- all such analysis happens at \textit{compile time}. Techniques include static type checking and type inference, data flow and information flow analysis, abstract interpretation, and symbolic execution. Static analysis can be extremely useful for automatically error checking code, as well as performing code optimisation.

This part focuses on static analysis for probabilistic programming. It starts with a short overview of the notation and methods used (\autoref{ch:pl-background}).
It then introduces the probabilistic programming language SlicStan --- a more compositional version of Stan --- 
and presents two novel techniques (Chapters~\ref{ch:slicstan} and \ref{ch:slicstan2}) that automatically refine inference by transforming a SlicStan program at compile time while preserving its semantics.

%% file: chapters/pl-background.tex
\Chapter{Formal treatment of programming languages}{Background and intuition} \label{ch:pl-background}

This short chapter contains a description of formal treatment of programming languages and introduces the notation used in \autoref{part:static}. It is predominantly meant as a simple reference for some standard notation in programming languages research.

\section{Formal syntax of programming languages}

Usually, programming languages are studied through an idealised calculus defined through a grammar of terminal and non-terminal symbols. As an example, let's consider a very simple arithmetic language defined by the following grammar, where $c$ ranges over numbers and $x$ ranges over strings: 

\begin{display}{Simple Arithmetic Language}
	\Category{E}{expression} \\
	\entry{c}{constant} \\
	\entry{x}{variable} \\
	\entry{E_1 + E_2}{addition} \\
	\entry{E_1 - E_2}{subtraction}
\end{display}

A program in this language is an expression $E$, which can be a constant, a variable, a sum of two expressions, or the difference between two expressions. For example, $x + 2$ and $3.0 - 2.1 + z + w$ are valid programs, while $2 +\!+ \;3$ and $2\;-$ are not.

\section{Rules of inference}

Aspects of programming languages, such as semantics or type systems, are often formalised through \textit{inductive definitions}. An inductive definition for some relation is the least relation (according to an underlying order), which is closed under some set of rules (for an example, refer to \citet{AndyTutorial95}).
One common way of making an inductive definition is through a set of \textit{inference rules}:
rules that consist of one or more \textit{premises} $P_n$ separated via a horizontal line from a \textit{conclusion} $Q$:
$$\copyrule{P_1 \quad \dots \quad P_N}{Q}$$
The above rule states that if we know that $P_1, \dots, P_N$ all hold then we can conclude $Q$. The rule is often read in the other direction, from the bottom to the top: we can conclude $Q$ if we can separately conclude each $P_n$ for $n = 1, \dots, N$.  

For example, consider a simplified propositional logic, where a logic formula $P$ consists of only conjunctions and disjunctions of other formul\ae:
\begin{display}{Simplified Propositional Logic:}
	\Category{P}{formula}\\
	\entry{\mathbf{T}}{true}\\
	\entry{\mathbf{F}}{false}\\
	\entry{P_1 \wedge P_2}{conjunction}\\
	\entry{P_1 \vee P_2}{disjunction}
\end{display}

We can inductively define a relation $\vdash P$, read as ``$P$ is provable'', through the following set of inference rules:
\begin{display}{Rules of the Simplified Propositional Logic:}
	\qquad
	\staterule{True}{}{\vdash \mathbf{T}} \qquad
	\staterule{And}{\vdash P_1 \qquad \vdash P_2}{\vdash P_1 \wedge P_2} \qquad
	\staterule{Or1}{\vdash P_1}{\vdash P_1 \vee P_2} \qquad
	\staterule{Or2}{\vdash P_2}{\vdash P_1 \vee P_2} \qquad
\end{display}

The rule \ref{True} is an \textit{axiom}: the formula $\mathbf{T}$ is provable without any premise.

On the other hand, \ref{And} states that if $P_1$ is provable and $P_2$ is provable, than we can conclude that $P_1 \wedge P_2$ is provable. Alternatively, as the term $\vdash P_1 \wedge P_2$ appears as a conclusion only in rule \ref{And}, we know that to show $P_1 \wedge P_2$ is provable we must show that $P_1$ is provable and that $P_2$ is provable. This is in contrast to disjunction, where to show $\vdash P_1 \vee P_2$ we must either show that $P_1$ is provable (by applying \ref{Or1}) or that $P_2$ is provable (by applying \ref{Or2}).

We can use these rules to check if a formula is provable by deriving a proof tree. For example:
\vspace{-8pt}
\begin{displaymath}
	\prftree[r]{$\scriptstyle\ref{And}$}
	{\prfbyaxiom{$\scriptstyle\ref{True}$}{\vdash \mathbf{T}}}
	{\prftree[r]{$\scriptstyle\ref{Or2}$}
		{\prftree[r]{$\scriptstyle\ref{Or1}$}
			{\prfbyaxiom{$\scriptstyle\ref{True}$}{\vdash \mathbf{T}}}
			{\vdash \mathbf{T} \vee \mathbf{F}}}
		{\vdash \left(\mathbf{T} \wedge \mathbf{F}\right) \vee \left(\mathbf{T} \vee \mathbf{F}\right)}}
	{\vdash \mathbf{T} \wedge \left[\left(\mathbf{T} \wedge \mathbf{F}\right) \vee \left(\mathbf{T} \vee \mathbf{F}\right)\right]}
\end{displaymath} 
\vspace{-24pt}

A proof tree of a formula that is not provable does not exist. For example, it is not possible to derive a proof tree of $\vdash \mathbf{F} \wedge \mathbf{T}$. If it was possible, it would be through the rule \ref{And}, meaning we would need to derive a proof tree for $\vdash \mathbf{T}$ and $\vdash \mathbf{F}$. Since there is no rule that can show $\vdash \mathbf{F}$, it is also not possible to show a proof for $\vdash \mathbf{F} \wedge \mathbf{T}$.

Inference rules allow for a concise formal definition of a calculus and can be used to derive proofs of various properties. This is often either by \textit{structural induction}, where we assume a property holds for the sub-terms building up an expression in the calculus, or by \textit{rule induction}, which is induction on the height of the proof tree for a property. Structural induction and rule induction are used in various proofs throughout Chapters~\ref{ch:slicstan} and \ref{ch:slicstan2}.

As an example of rule induction, let us prove a simple property of the simplified logic of this section: if a formula is provable then it must contain $\mathbf{T}$ as a subterm. 

To state this formally, we introduce a new relation $\mathrm{t}(P)$, read ``$P$ contains $\mathbf{T}$'', defined as:
\vspace{-24pt}
\begin{display}{}
	\qquad
	\staterule{T True}{}{\mathrm{t}(\mathbf{T})} \qquad
		\staterule{T And1}{\mathrm{t}(P_1)}{\mathrm{t}(P_1 \wedge P_2)} \qquad
	\staterule{T And2}{\mathrm{t}(P_2)}{\mathrm{t}(P_1 \wedge P_2)} \qquad
	\staterule{T Or1}{\mathrm{t}(P_1)}{\mathrm{t}(P_1 \vee P_2)} \qquad
	\staterule{T Or2}{\mathrm{t}(P_2)}{\mathrm{t}(P_1 \vee P_2)} \qquad
\end{display}
\vspace{-2pt}
\begin{proposition}[Provable formul\ae~contain the subterm $\mathbf{T}$] ~\\
	For any formula $P$, $\vdash P$ implies $\mathrm{t}(P)$.
\end{proposition}
\vspace{-4pt}
\begin{proof}
	We prove by rule induction on the derivation of $\vdash P$. 
	
	We start by the base case when the proof tree for $\vdash P$ is of height 1. In other words, the whole proof tree must be the axiom \ref{True}:
	\begin{itemize}
	\item Case \ref{True}. $P = \mathbf{T}$. Thus $\mathrm{t}(P)$ by \ref{T True}.
	\end{itemize} 

	This concludes the induction's base case. Next, we assume the induction hypothesis that for any formula $P'$ such that $\vdash P'$, and the height of the proof tree of $\vdash P'$ is smaller than or equal to $n$, we have $\mathrm{t}(P')$. Consider a formula $P$ with $\vdash P$ proof tree of height $n+1$. Then, it must be the case that we derived $\vdash P$ using one or more proof tree of height $ \leq n$ in combination with one of the following rules:
	\begin{itemize}
	\item Case \ref{And}. $P = P_1 \wedge P_2$, for some $P_1$ and $P_2$. According to \ref{And}, it must be the case that $\vdash P_1$ and $\vdash P_2$. But the proof tree of each of those is of height at most $n$, thus, by the induction hypothesis, $\mathrm{t}(P_1)$ and $\mathrm{t}(P_2)$. By \ref{T And1} (or \ref{T And2}), it follows that $\mathrm{t}(P)$.
	\item Case \ref{Or1}. $P = P_1 \vee P_2$, for some $P_1$ and $P_2$. According to \ref{Or1}, it must be the case that $\vdash P_1$. But the proof tree for $\vdash P_1$ is of height at most $n$, thus, by the induction hypothesis, $\mathrm{t}(P_1)$. By \ref{T Or1} it follows that $\mathrm{t}(P)$.	
	\item Case \ref{Or2}. $P = P_1 \vee P_2$, for some $P_1$ and $P_2$. According to \ref{Or2}, it must be the case that $\vdash P_2$. But the proof tree for $\vdash P_2$ is of height at most $n$, thus, by the induction hypothesis, $\mathrm{t}(P_2)$. By \ref{T Or2} it follows that $\mathrm{t}(P)$.	
\end{itemize} 
	This concludes the induction step and hence the proof.
\end{proof}

\section{Semantics of programming languages}

The semantics of a programming language is a formal specification of its meaning. This can be done by formalising the execution of a program (\textit{operational semantics}, \citet[Chapter~2]{Plotkin1981, Nielson92}) or by attaching a mathematical meaning to the terms of the language (\textit{denotational semantics}, \citet[Chapter~5]{Nielson92}). As an example, let's give the operational semantics of our arithmetic language. 

Suppose $\sigma$ is a mapping from variable names $x_n$ to concrete values $V_n$: $\sigma = \{x_1 \mapsto V_1, \dots x_N \mapsto V_N\}$. The semantics of the language is given by the relation $(\sigma, E) \Downarrow V$, which is such that an expression $E$, where variables are substituted with values according to $\sigma$, computes a value $V$. We can read $(\sigma, E) \Downarrow V$ as ``In the context of $\sigma$ expression $E$ evaluates to $V$.'' The operational semantics can be specified recursively for different terms of the calculus as follows:  

\begin{display}[.50]{Operational Semantics of the Arithmetic Language:}
	\quad
	\staterule{Eval Cst}{}{(\sigma, c) \Downarrow c} \quad
	\staterule{Eval Var}{x \in \dom(\sigma)}{(\sigma, x) \Downarrow \sigma(x)} \quad
	\staterule{Eval Add}{(\sigma, E_1) \Downarrow V_1 \quad (\sigma, E_2) \Downarrow V_2}{(\sigma, E_1 + E_2) \Downarrow V_1 + V_2} \quad
	\staterule{Eval Sub}{(\sigma, E_1) \Downarrow V_1 \quad (\sigma, E_2) \Downarrow V_2}{(\sigma, E_1 - E_2) \Downarrow V_1 - V_2} \quad
\end{display}

\ref{Eval Cst} simply states that a constant $c$ always evaluates to the same constant $c$, regardless of the store $\sigma$. On the other hand, as per \ref{Eval Var} a variable $x$ evaluates to the value of $x$ given by $\sigma$, as long as $x$ is in the domain of $\sigma$. \ref{Eval Add} states that to evaluate $E_1 + E_2$ we must evaluate $E_1$ and $E_2$ separately and add the result. Here, the usage of $+$ in $V_1 + V_2$ is understood as standard addition between values and not a symbol of the calculus.  
Similarly, \ref{Eval Sub} requires evaluating the subexpressions $E_1$ and $E_2$ to evaluate $E_1 - E_2$.

\section{Type checking and type inference}

Type checking \cite[Chapter~5]{Pierce02, Nielson04} is a form of program analysis where we verify the correct usage of types, such as integer, boolean, array or functional types. Static type checking analyses the parse tree of a program without running it, to flag any type discrepancies. This is useful, as it allows us to detect possible mistakes in the code at compile time. 

To type-check programs in some language, we need to formalise its \textit{type system}, which includes rules for building types and typed expressions in the language. For example, let's define the type system of our arithmetic language as follows:

\begin{multicols}{2}
\begin{display}[.50]{Types of the Arithmetic Language:}
	\clause{\tau ::= \kw{real} \mid \kw{int}}{} 
\end{display}

\begin{display}[.50]{Typing Environment:}
	\clause{\Gamma ::= \{x_1\mapsto \tau_1, \dots, x_n\mapsto \tau_n\}}{}
\end{display} 
\end{multicols}

\begin{display}[.35]{Judgement of the Type System:}
	\clause{\Gamma \vdash E : \tau}{expression $E$ has type $\tau$ according to $\Gamma$}
\end{display} 

A type in the language is denoted by $\tau$, which is either $\kw{real}$ --- a floating point number --- or $\kw{int}$ --- an integer. The typing environment holds information about the types of variables in the form of a mapping from variable names to types. 
The typing judgement $\Gamma \vdash E : \tau$ is a relation on $\Gamma, E$ and $\tau$ such that an expression $E$ is of type $\tau$ if the types of variables in $E$ are given by $\Gamma$. 
The typing rules specify what programs in the language are \textit{well-typed}: what is needed for the relation $\Gamma \vdash E : \tau$ to hold. 

\newpage
\begin{display}[.50]{Typing Rules of the Arithmetic Language:}
	\quad
	\staterule{Cst}{\tau = \kw{ty}(c)}{\Gamma \vdash c : \tau} \hquad\qquad
	\staterule{Var}{\Gamma(x) = \tau}{\Gamma \vdash x : \tau} \hquad\qquad
	\staterule{Add}{\Gamma \vdash E_1 : \tau \quad \Gamma \vdash E_2 : \tau}{\Gamma \vdash E_1 + E_2 : \tau} \hquad\qquad
	\staterule{Sub}{\Gamma \vdash E_1 : \tau \quad \Gamma \vdash E_2 : \tau}{\Gamma \vdash E_1 - E_2 : \tau} \hquad\quad
\end{display}

\ref{Cst} simply states that the program type of a constant $c$ corresponds to its numerical type (we assume $\kw{ty}(c) = \kw{int}$ if $c \in \mathbb{Z}$ and $\kw{ty}(c) = \kw{real}$ if $c \in \mathbb{R}$ and $c \notin \mathbb{Z}$). \ref{Var} states that the type of a variable is determined by the typing environment $\Gamma$. \ref{Add} and \ref{Sub} define well-typedness of expressions of the form $E_1 \pm E_2$: in order for $E_1 \pm E_2$ to be well-typed of type $\tau$ in $\Gamma$, it needs to be the case that both $E_1$ and $E_2$ are well-typed of type $\tau$ in $\Gamma$. This means that $1.0 + (x - 2.5)$ is well-typed in $\Gamma = \{x \mapsto \kw{real}\}$, but it is not well-typed in $\Gamma = \{x \mapsto \kw{int}\}$.

Some typed languages require all variable types to be explicitly specified as part of the program and perform type checking as a form of a correctness check. It is also possible to work with only partially specified types and \textit{infer} any missing types through \textit{type inference} \cite[Chapter~22]{Pierce02}. For example, by analysing the program $1.0 + (x - 2.5)$ according to the typing rules of our arithmetic language, we can deduce that $x$ must be of type $\kw{real}$ in order for the program to be well-typed.

\section{Information flow analysis} \label{sec:bg-infoflow}

One program analysis technique central to the contributions of \autoref{part:static} is \textit{information flow analysis} \cite{Volpano96, Abadi99}.
Information flow is the transfer of information between two variables in a computation. 
In the program $y = x + 1$ information flows from $x$ to $y$. 

Analysing the flow of information can be especially useful in detecting program vulnerabilities.
Static analysis techniques concerning the flow of information are popular in the security community, where the goal is to prove that systems do not leak information.
Secure information flow analysis, summarised by \citet{Sabelfeld2003}, and \citet{Smith2007}, concerns systems where variables have one of several security levels. 

For example, suppose we have two security levels, \lev{low} and \lev{high}. Variables of level \lev{low} are \textit{low security}; they hold \textit{public data}. In contrast, variables of level \lev{high} are of \textit{high security}; they hold \textit{secret data}.
In such system, we want to disallow the flow of secret information to a public variable, but allow other flows of information. That is, if $L$ is a variable of security level \lev{low}, $H$ is a variable of security level \lev{high}, and $f$ is some function, we want to forbid statements such as $L = f(H)$, but allow:
\begin{itemize}
	\item  $L = f(L)$
	\item  $H = f(H)$
	\item  $H = f(L)$
\end{itemize}

But there are ways other than simple data flow, in which information can leak to a lower security level. For example through conditional statements. Suppose $L$ and $H$ from above are boolean variables. We want to disallow statements such as $\kw{if}\;\!(H)\;\kw{then}\;L = \kw{true}$, as this leaks information about the higher security level $H$ to the lower security level $L$.

In the general case, we may be interested in having more than two information levels.
Formally, those levels form a \textit{lattice} --- a partially ordered set $(L;<)$, where every two elements of $L$ have a unique least upper bound and a unique greatest lower bound.
Secure information flow analysis is used to ensure that information flows only upwards with respect to that lattice. In the case with \lev{low} and \lev{high}, information flows only from \lev{low} to \lev{high} and never in the other direction.   
This is also known as the \textit{noninterference} property --- changes to confidential inputs lead to no changes in public outputs of a system \cite{Goguen1982}. 

One way to ensure noninterference is by formalising a type system such that each variable has a level type and noninterference holds for any well-typed program. This is also the approach used in this dissertation, as discussed in Chapters~\ref{ch:slicstan} and \ref{ch:slicstan2}.

%% file: chapters/slicstan1.tex
\cleardoublepage
\Chapter{SlicStan}{Optimising pre-and post-processing code for inference} \label{ch:slicstan}

\vspace{-6pt}
This chapter describes SlicStan: a version of Stan that contains no \textit{program blocks} and is more compositional. 
The main function of Stan's program blocks is to explicitly separate the program into a \textit{pre-processing} part (executed only once), \textit{post-processing} part (executed after each sample is drawn), and the \textit{core} of the model, which is executed to generate a single sample and requires the most computational resources. SlicStan demonstrates that such separation need not be done manually by the programmer, but can be automated using \textit{information-flow analysis}.
We proceed with the main contribution of the chapter, 
the paper \textit{Probabilistic Programming with Densities in SlicStan: Efficient, Flexible, and Deterministic} (\autoref{sec:slicstan-paper}), clarifying the contributions with respect to previous work (\autoref{sec:slicstan-contribs}), and discussing the impact of SlicStan (\autoref{sec:slicstan-impact}). 

\vspace{-3pt}
\section{The paper} \label{sec:slicstan-paper}
\vspace{-1pt}

This section presents the work \textit{Probabilistic Programming with Densities in SlicStan: Efficient, Flexible, and Deterministic}. The paper gives the first formal semantics of Stan and introduces SlicStan: a more compositional version of Stan. It describes a semantic-preserving procedure for translating SlicStan to Stan. 
By combining information-flow analysis and type inference, SlicStan performs automatic program optimisation, which frees users from the need to explicitly encode the parts of the program that correspond to pre-processing, post-processing and the computationally heavy core of the model.

The paper was accepted for presentation at the \textit{46th ACM SIGPLAN Symposium on Principles of Programming Languages (POPL 2019)} and included in the \textit{Proceedings of the ACM on Programming Languages, Volume 3, Issue POPL}. Out of 267 papers submitted in total, 77 papers were accepted.  

\newpage
\includepdf[pages=-,addtotoc={
	1, subsection, 1, Introduction, p1,
	1, subsubsection, 1, Background: Probabilistic Programming Languages and Stan, p1,
	2, subsubsection, 1, Goals and Key Insight, p2,
	2, subsubsection, 1, The Insight by Example, p2,
	3, subsubsection, 1, Core Contributions and Outline, p3,
	3, subsection, 1, Core Stan, p3,
	3, subsubsection, 1, Syntax of Core Stan Expressions and Statements, p3,
	4, subsubsection, 1, Operational Semantics of Stan Statements, p4,
	5, subsubsection, 1, Syntax of Stan, p5,
	6, subsubsection, 1, Density-Based Semantics of Stan, p6,
	8, subsubsection, 1, Inference, p8,
	8, subsection, 1, SlicStan, p8,
	9, subsubsection, 1, Syntax, p9,
	10, subsubsection, 1, Typing of SlicStan, p10,
	13, subsubsection, 1, Elaboration of SlicStan, p13,
	15, subsubsection, 1, Semantics of SlicStan, p15,
	15, subsubsection, 1, Examples, p15,
	17, subsubsection, 1, Difficulty of Specifying Direct Semantics Without Elaboration, p17,
	17, subsection, 1, Translation of SlicStan to Stan, p17,
	17, subsubsection, 1, Shredding, p17,
	19, subsubsection, 1, Transformation, p19,
	21, subsection, 1, Examples and Discussion, p21,
	21, subsubsection, 1, Type Inference, p21,
	22, subsubsection, 1, Locality, p22,
	23, subsubsection, 1, Code Refactoring, p23,
	24, subsubsection, 1, Code Reuse, p24,
	25, subsection, 1, Related Work, p25,
	26, subsubsection, 1, Formalisation of Probabilistic Programming Languages, p26,
	26, subsubsection, 1, Static Analysis for Probabilistic Programming Languages, p26,
	27, subsubsection, 1, Usability of Probabilistic Programming Languages, p27,
	27, subsection, 1, Conclusion, p27,
	31, subsection, 1, Appendix A: Definitions and Proofs, p31,
	31, subsubsection, 1, Definitions, p31,
	32, subsubsection, 1, Proof of Semantic Preservation of Shredding, p32,
	37, subsection, 1, Appendix B:Further discussion on semantics, p37,
	37, subsubsection, 1, Semantics of Generated Quantities, p37,
	38, subsubsection, 1, Relation of Density-based Semantics to Sampling-based Semantics, p38,
	39, subsection, 1, Appendix C: Elaborating and shredding if or for statements, p39,
	40, subsection, 1, Appendix D: Non-centred Reparameterisation, p40,
	41, subsection, 1, Appendix E: Examples, p41,
	41, subsubsection, 1, Neal's Funnel, p41,
	45, subsubsection, 1, Cockroaches, p45,
	47, subsubsection, 1, Seeds, p47
}]{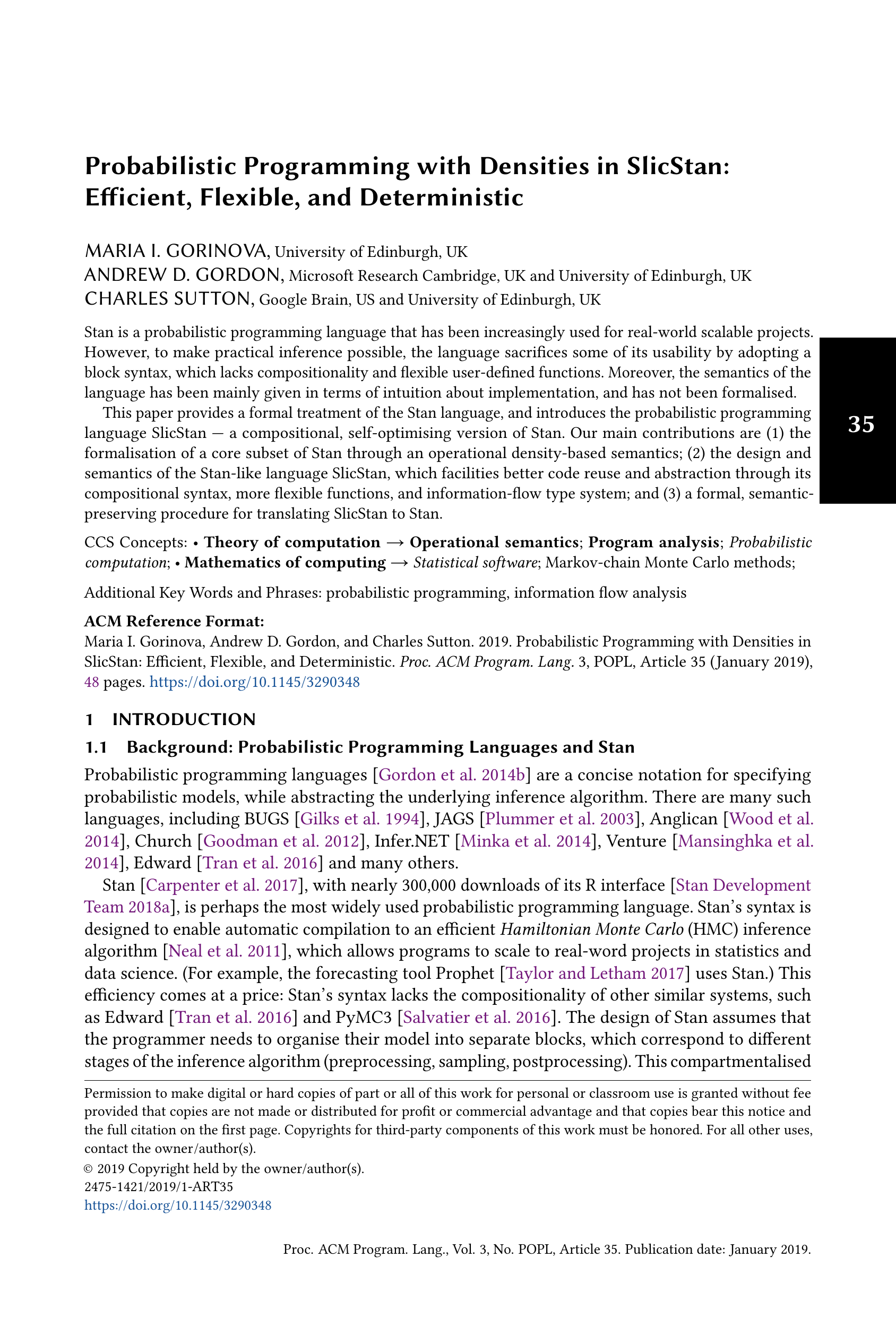}
\newpage

\section{Clarifying the contributions} \label{sec:slicstan-contribs}

\paragraph{Author contributions.} The authors of the paper are myself, Andy Gordon and Charles Sutton. Andy Gordon conceived the idea about using information flow analysis to slice a SlicStan program into Stan blocks and suggested using operational semantics to formalise Stan and SlicStan. Both Andy Gordon and Charles Sutton supervised the project throughout, gave comments and suggestions, and helped write and revise the paper. As a leading author, my contribution included working out the details of the Stan/SlicStan semantics and type system, implementing SlicStan, deriving the proofs in the paper, preparing code examples, and co-writing the paper.

\paragraph{Similarities to previous work.} The SlicStan project started before the onset of this PhD program. Part of the work presented in the paper, and thus in this chapter, was also presented in the \textit{Probabilistic Programming with SlicStan} MSc dissertation \cite{SlicStanMSc}.
In particular, the MSc dissertation included an initial version of SlicStan's syntax and information-flow type system ($\S\S~3.1$ and $\S\S~3.2$ of the paper), as well as an elaboration and transformation procedures for this reduced version of SlicStan ($\S\S~3.3$ and $\S\S~4.2$ of the paper). While similar in spirit to the final paper, these parts of the MSc dissertation concerned only a subset of SlicStan's final syntax and were significantly developed, as clarified below. 
Finally, the comparison between Stan and SlicStan ($\S~5$ and Appendix E from the paper) was largely adapted from the MSc dissertation.

\paragraph{Contributions of this chapter.} The SlicStan paper, and thus this chapter, contain the following changes and contributions that are new compared to the MSc dissertation:
\begin{itemize}
\item The syntax of SlicStan was extended to include conditional statements and \kw{for}-loops. The new syntax also omits having separate symbol for distributions, making the calculus more concise. It treats terms such as $E \sim d(E_1, \dots, E_n)$ as derived forms rather than as part of the core calculus. (Although, the last change was reversed in \autoref{ch:slicstan2}.)
\item The type system presented in the MSc dissertation contained an oversight, which made it difficult to correctly slice SlicStan programs such as:
\begin{lstlisting}
		real x = 0;
		real y ~ normal(x, 1);
		x = 1;
\end{lstlisting}
In the initial, MSc dissertation version of SlicStan, this program would type-check when $x$ is of level \lev{data} and $y$ is of level \lev{model}. However, such type level assignment would result in the following Stan program:
\begin{lstlisting}
	 	transformed data { 
	 		real x = 0; 
 			x = 1;
 		}
 		parameters { real y; }
 		model { y ~ normal(x, 1); }
\end{lstlisting}\vspace{-4pt}
As blocks in Stan are executed in order, the meaning of the program has changed in the translation: the statement \lstinline{x = 1} must appear after \lstinline{y ~ normal(x, 1)}, but appears before it. This is not a problem of the translation procedure, but a type problem. It is not possible for $x$ to be of level \lev{data} and for the statements to be in the correct order, as any transformation of a variable of level \lev{data} must appear in the \kw{transformed parameters} block, which appears strictly before \lev{model}-level statements such as \lstinline{y ~ normal(x, 1)}.

This chapter corrects the above problem by ensuring that such programs fail to type-check. It introduces the concept of \textit{shreddable sequence} (Definition~$4.7$ of the paper), so that variables become immutable once they are read at a higher level than their own. In the paper, the program above fails to type-check if $x$ is of level \lev{data} and $y$ is of level \lev{model}, but type-checks for $x$ and $y$ both of level \lev{model}.

\item Extending SlicStan with conditionals and loops required a more careful look at the rules for translating SlicStan to Stan. For example, conditional statements that contain sub-statements at several levels, need to be sliced by copying the guard of the \kw{if}-statement per sub-statement level: 
%
\vspace{-6pt}
\begin{multicols}{2}
\textbf{SlicStan} \vspace{-5pt}
\begin{lstlisting}
	data bool g;
	data real x;
	model real y;
	if (g) {
		x = 1;
		y ~ normal(x, 1);
	} 
	else {
		x = -1;
		y ~ normal(x, 2);
	}
\end{lstlisting} 

\textbf{Stan} \vspace{-5pt}
\begin{lstlisting}
	data { bool g; }
	transformed data {
		real x;
		if (g) { x = 1;	}
		else { x = -1; }
	}
	parameters { real y; }
	model {
		if (g) { y ~ normal(x, 1); }
		else { y ~ normal(x, 2); }
	}
\end{lstlisting}
\end{multicols}

Such slicing was formalised through the new to the paper \textit{shredding relation} ($\S\S~4.1$).

\item The biggest contribution of the paper presented in this chapter that is new compared to the MSc thesis is the formalised density-based semantics of Stan and SlicStan. This includes most of $\S~2$ and $\S~3.4$, which are entirely new to the paper. 

\item Additionally, the paper utilises this formal semantics by stating and proving several key results, most notably the semantic preservation of the transformation that translates SlicStan programs to Stan programs (Theorem~$4.10$ of the paper).
\end{itemize}

\vspace{-7pt}
\section{Impact} \label{sec:slicstan-impact}
\vspace{-4pt}
The idea of a ``blockless Stan'' and using information-flow to achieve it has influenced several lines of work. \citet{Yaps} describe a new probabilistic programming language, Yaps, which is inspired by SlicStan to allow for a concise Python-based frontend for Stan. 
In addition, the density-based semantics of Stan that the paper presents has been used as a basis to formalise a procedure for translating Stan to generative (implicit) PPLs   \cite{Baudart2021CompilingST}.
AQUA \cite{AQUA} is a PPL that uses symbolic inference and quantization of the probability density, whose semantics has also been inspired by that of SlicStan.
The SlicStan paper is also one of the first to utilise static analysis for probabilistic programming to improve inference, as summarised by \citet{BernsteinStatic19}.

While the paper presented here treats the program slicing as a way to divide the program into pre-inference, post-inference and inference, there is an alternative interpretation of the three slices. In Stan, the post-processing code, which is placed in the \kw{generated quantities} block, can include calls to pseudo-random number generation functions. The program as a whole can then be seen as inferring some variables (those defined in the \kw{parameters} block) via HMC, and others (those drawn with pseudo-random number generators) via ancestral sampling. This idea is further discussed and developed in the next chapter. 

Notably, SlicStan has been a point of discussion when it comes to the future of Stan. The \textit{\StanPP/Stan3 Preliminary Design} discussion \cite{Stan3Plans} features SlicStan and suggests adapting the information-flow approach to allow for more compositional feature version of Stan. However, this plan has not been made official or realised. 

\vspace{-7pt}
\section*{Errata}
\vspace{-4pt}

The example on page 17 of the paper is a valid SlicStan program that compiles. However, running it would result in a runtime error, as the variable \lstinline{d} is used before it is defined.

The program on page 24 has both variables \lstinline{x_mean} and \lstinline{x_meas}. Read both as \lstinline{x_mean}.

%% file: chapters/slicstan2.tex
\cleardoublepage
\Chapter{Conditional independence by typing}{Variable elimination for inference with discrete parameters} \label{ch:slicstan2}

Stan has often been criticised for its lack of \textit{explicit} support for discrete parameters, which is a consequence of using gradient-based algorithms for inference. 
However, a workaround is possible, where the user can manually sum out any discrete parameters, thus encoding them in an \textit{implicit} way. 
This chapter builds upon the \textit{information-flow analysis} ideas of \autoref{ch:slicstan} and presents a type system for conditional independence analysis, which can be used to automate the process of discrete variable marginalisation in SlicStan. 
The main contribution of the chapter is the paper \textit{Conditional Independence by Typing} (\autoref{sec:slicstan2-paper}). While focused on conditional independence and variable elimination, this work can be seen as an instance of a more general framework for static analysis of probabilistic programs, which I briefly discuss in \autoref{sec:slicstan2-discuss}.

\section{The paper} \label{sec:slicstan2-paper}

This section presents the work \textit{Conditional Independence by Typing}. 
The paper builds upon the original SlicStan work and presents an information-flow type system that captures conditional independence relationships in probabilistic programs. 
The paper shows how certain conditional independence relationships can be automatically deduced using type-inference. It presents one practical application of the type system: a semantic-preserving transformation, which can transform SlicStan programs with discrete parameters in a way that allows for efficient gradient-based inference. 

The paper was accepted for publication at the \textit{ACM Transactions on Programming Languages and Systems}, Volume 44, Issue 1 \textit{(TOPLAS 2022)}.

\paragraph{Author contributions.}
The paper is co-authored by me, Andy Gordon, Charles Sutton and Matthijs V\'ak\'ar. As a leading author, my contributions included conceiving the idea of using information-flow for conditional independence analysis, working out the details of the conditional independence type system and the program transformation for marginalising discrete variables, implementing the analysis within SlicStan, deriving some of the proofs in the paper, preparing and performing the experiments, and co-writing the paper. Both Andy Gordon and Charles Sutton supervised the project throughout, gave comments and suggestions, and co-wrote and revised the paper. Matthijs V\'ak\'ar contributed to adapting SlicStan's semantics and typing rules to cover generated quantities, helped derive some of the definitions and proofs, and co-wrote parts of the paper.

\newpage
\includepdf[pages=-,addtotoc={
	1, subsection, 1, Introduction, p1,
	2, subsection, 1, SlicStan: Extended syntax and semantics, p2,
	4, subsubsection, 1, Syntax, p4,
	5, subsubsection, 1, Typing, p5,
	6, subsubsection, 1, Operational Semantics of SlicStan Statements, p6,
	9, subsubsection, 1, Density Semantics, p9,
	10, subsubsection, 1, Shredding and Translation to Stan, p10,
	12, subsubsection, 1, Density Factorisation, p12,
	12, subsection, 1, Theory: Conditional independence by typing, p12,
	15, subsubsection, 1, The $\vdash _{2}$ Type System, p15,
	18, subsubsection, 1, Conditional Independence Result for $\vdash _{2}$-Well-Typed Programs, p18,
	18, subsubsection, 1, Scope of the Conditional Independence Result, p18,
	19, subsection, 1, Application: Discrete parameters support through a semantics-preserving transformation, p19,
	20, subsubsection, 1, Goal, p20,
	21, subsubsection, 1, Key Insight, p21,
	22, subsubsection, 1, Variable Elimination, p22,
	22, subsubsection, 1, Conditional Independence and Inferring the Markov Blanket, p22,
	23, subsubsection, 1, Sampling the Discrete Parameters, p23,
	24, subsubsection, 1, A Semantics-Preserving Transformation Rule, p24,
	26, subsubsection, 1, Marginalising Multiple Variables: An example, p26,
	30, subsubsection, 1, Relating to Variable Elimination and Complexity Analysis, p30,
	30, subsubsection, 1, Semantic Preservation of the Discrete Variable Transformation, p30,
	31, subsubsection, 1, Scope and limitations of {\small \sc Elim Gen}, p31,
	33, subsection, 1, Implementation and empirical evaluation, p33,
	33, subsubsection, 1, Implementation, p33,
	33, subsubsection, 1, Empirical evaluation, p33,
	37, subsubsection, 1, Analysis and discussion, p37,
	38, subsection, 1, Related work, p38,
	39, subsection, 1, Conclusion, p39,
	42, subsection, 1, Appendix A: Definitions and Proofs, p42,
	42, subsubsection, 1, Definitions, p42,
	43, subsubsection, 1, Proofs, p43,
	51, subsection, 1, Appendix B: Examples, p51,
	51, subsubsection, 1, Sprinkler, p51,
	51, subsubsection, 1, Soft-K-means model, p51,
	52, subsubsection, 1, A causal inference example, p52
}]{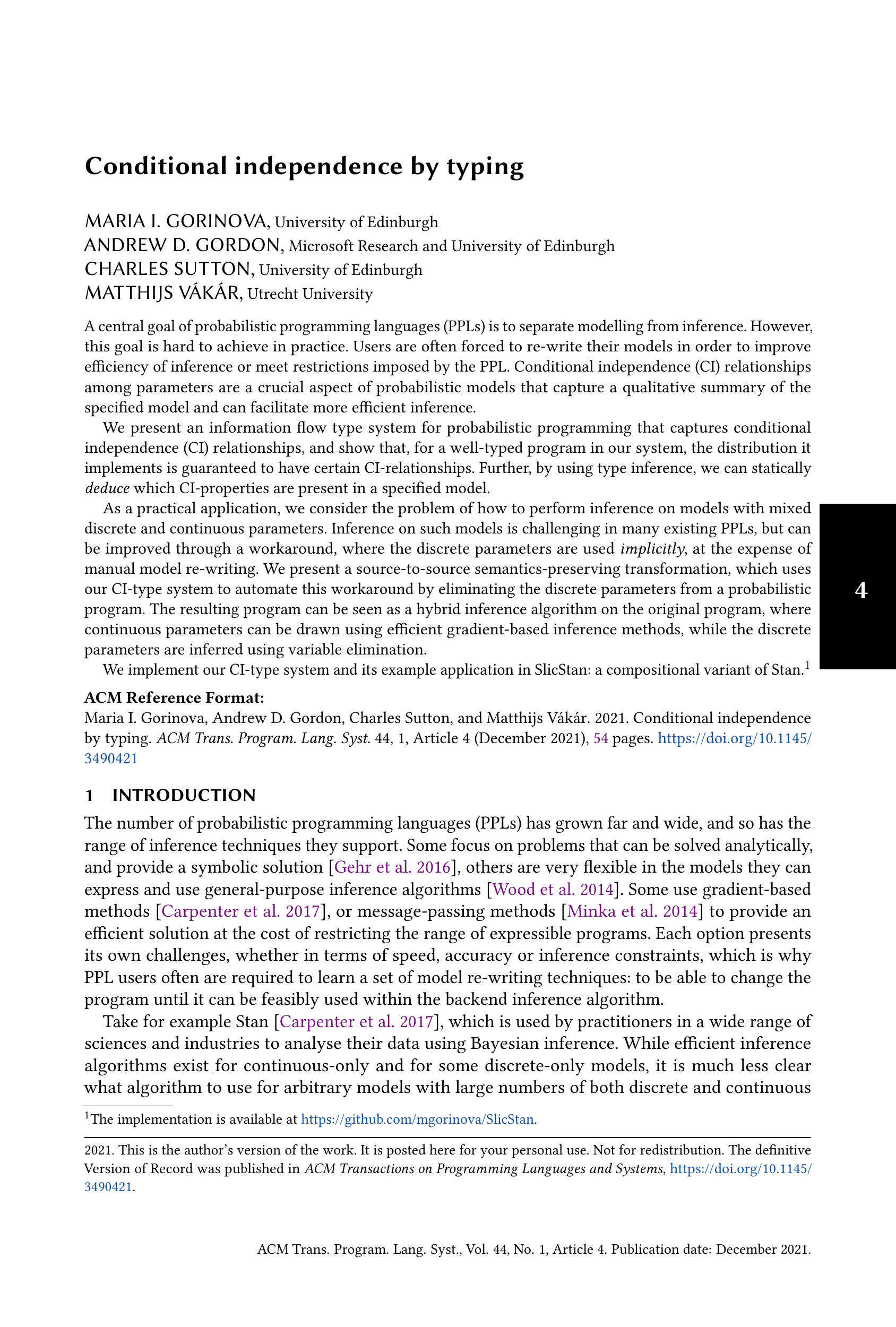}
\newpage

\section{Discussion} \label{sec:slicstan2-discuss}

The paper presents two type systems:
\begin{enumerate}
	\item \textbf{Generative subprogram type system $\vdash$.}~ The first type system is close to the original SlicStan (\autoref{ch:slicstan}), where code is split into pre-processing, core, and post-processing parts. The improvement implemented in the current chapter allows us to treat sampling statements as either contributing to the evaluation of the target density function (density-based semantics) or as literal sampling (sampling-based semantics) based on context. This allows for a SlicStan program to be sliced into a \textit{pre-processing} part, \textit{core} part, where inference is done using Hamiltonian Monte Carlo or another resource-consuming inference algorithm, and purely \textit{generative} part, where samples can be generated through ancestral sampling. That is, the sliced program corresponds to \textit{HMC-AS hybrid inference}.
	\item \textbf{Conditional independence type system $\typingspecial$.}~ The second type system is the main contribution of the paper and allows for exploring conditional independence relationships. One way in which this system can be used is to transform a program with discrete parameters by marginalising out variables efficiently. 
	The transformed program can be seen as a hybrid inference algorithm on the original program, where continuous parameters are drawn using Hamiltonian Monte Carlo or another gradient-based algorithm, while the discrete parameters are drawn using variable elimination. That is, the transformed program corresponds to \textit{HMC-VE hybrid inference}. 
\end{enumerate}

While different, the two type systems, and the program slicing associated with each, share a lot of elements. For example, noninterference holds for both systems (Lemma 1 and Lemma 7). The two shredding relations associated with each of $\vdash$ and $\typingspecial$ are identical, up to the naming of level types. Shredding a SlicStan statement with respect to either type system produces single level statements (Lemma 4 and Lemma 9), preserves semantics (Lemma 6 and 10), and induces a (type system dependent) factorisation of the density (Theorem 1 and Theorem 2).

This hints at the possibility of generalising the approach.
In particular, future work may build a general framework for static analysis of probabilistic programs, where the choice of partially ordered set and the design of typing rules can be motivated by a particular factorisation of interest.
If such a framework can be built in a modular way, it could serve as a basis for a PPL that supports composable and programmable inference.

\section*{Errata}

The sentence ``Sometimes, when $\sigma$ is
clear from context, we will leave it implicit and simply write $p(\mathbf{x})$ for $p(\mathbf{x}; \sigma)$.'' on page 4:9 of the paper is an artefact left from previous version of the notation and is redundant.

%% file: chapters/part2-intro.tex
\section*{Introduction to Part II}

In the previous part, we saw examples of \textit{static} analysis for probabilistic programming, that is analysis that is performed at compile-time, before the program is executed. Such analysis can be incredibly useful for exploiting conditional independencies and performing semantic-preserving rewrites, which improve the performance of the underlying inference algorithm. However, inference performance is often affected not only by the structure of the probabilistic model in question, but also by the nature of the observed data.

\textit{Dynamic analysis} refers to the class of program analysis techniques that are performed \textit{during the execution} of a program.   
This part focuses on dynamic analysis for probabilistic programs and presents one such dynamic method in \autoref{ch:autoreparam}. The method transforms a program at run-time, taking into account both the model and the data, in order to improve the quality of inference results. 

%% file: chapters/autoreparam.tex
\cleardoublepage
\Chapter{Automatic reparameterisation}{Variational inference pre-processing for efficient inference}
\label{ch:autoreparam}

This chapter presents an instance of dynamic program analysis for probabilistic programming: an automatic model reparameterisation procedure.
The \textit{parameterisation} of a model is the particular way in which the model is expressed in terms of parameters. What parameterisation we choose to express our model plays a vital role in the way the geometry of the posterior of interest looks like, and hence (as we saw in \autoref{sec:bg-inference}), has a direct effect on the quality of inference.  
While changing the parameterisation of a model, or \textit{reparameterising} it, can be done in many ways (for example through a simple static transformation), this chapter looks at reparameterisation through the lens of effect-handling based PPLs. Such PPLs provide a dynamic way to interpret random assignment statements and I give further background on the topic in  \autoref{sec:autoreparam-eff-ppls}.
The main contribution of the chapter is the paper
\textit{Automatic Reparameterisation of Probabilistic Programs} (\autoref{sec:autoreparam-paper}),
which presents a way to reparameterise Edward2 programs using effect-handlers and based on both the defined model and the observed data.
Finally, the chapter discusses the impact and limitations of the proposed approach (\autoref{sec:autoreparam-discuss}).

\section{Effect-handling based probabilistic programming} \label{sec:autoreparam-eff-ppls}

\autoref{ch:background} proposed classifying PPLs in three categories --- explicit (\autoref{ssec:bg-density-ppls}), implicit (\autoref{ssec:bg-trace-ppls}), and effect-handling based (\autoref{ssec:bg-eff-ppls}) --- and briefly introduced each category. In this section, I give further background on the mechanisms behind the last, effect-handling based type of probabilistic programming languages. This section is partially based on \citet{ProbProg18}.

\subsection{Effects and effect handling}

When working with complex code, or when we want to rigorously reason about the behaviour of a program, it is often useful to write \textit{pure} functions.
That is, deterministic functions that do not mutate a global state, or any of the input arguments, do not throw exceptions and do not perform I/O operations. They do not interact with the external world, or code outside of the function, in any way. Such functions have no \textit{side effects}.
While it is easier to reason about pure functions, it can also be very constraining in real-world situations.

\textit{Algebraic effects} and their \textit{handlers} have emerged as a convenient way to control \textit{impure} behaviour. 
They are built upon a strong mathematical foundation \cite{Effects, Effects2}, 
their semantics can be precisely defined \cite{Kammar2013}, 
and in some cases they provide an improvement in runtime complexity compared to pure languages \cite{EffectsEfficiency}.
While providing a useful formal framework for reasoning about side effects, algebraic effects and handlers have also proved to be an increasingly convenient modular abstraction that has been adopted across many disciplines. Some examples include concurrent programming \cite{Dolan2017}, meta programming \cite{Yallop2017}, and probabilistic programming \cite{Pyro}.

The premise behind algebraic effects and handlers is that impure behaviour arises from a set of \textit{effectful operations}: for example, \kw{read}, \kw{print}, \kw{set}, \kw{raise}, and so on. Such operations interact with some \textit{handling} code in order to execute. 
For example, consider a process that wants to \kw{read} from a file. It sends a 
request to the OS kernel and suspends execution. The kernel checks the request, executes it, and 
responds with the result of the \kw{read} operation. The process then resumes execution.
This idea of effect handlers as operating systems in further explored by \citet{HillerstromPhDThesis}.

While in the case of accessing a file, the handling code is external to the program, in some cases it could also be internal: for example, when an exception is risen, it can be \textit{handled} inside the same program. 
Algebraic effects and handlers extend this idea of handling programmatically exceptions, to any effectful operation. The concrete implementation of operations is given and managed by code written by the user. 

To demonstrate the workings of effect handlers in this section, Consider a simple example of a program that reverses the order of print statements given by \citet{EffectsTutorial}:
\begin{lstlisting}
	let abc = (print("a"); print("b"); print("c"))
	let reverse = handler { print(s; x. k) $\rightarrow$ k(); print(s) }
	with reverse handle abc
\end{lstlisting}
Here, \lstinline{print} is an \textit{operation}, while 
\lstinline{abc} is a \textit{computation} that prints out the letters `a', `b' and `c', in this order, using three separate calls to \lstinline{print}. 
Key to working with algebraic effects is the presence of \textit{continuations}: functions that specify how the program ``continues'' after a certain point; they express ``what to do next'' \cite{ContinuationsBook}. 
Thus, the program \lstinline{print("a")$\!$;$\,$print("b")$\!$;$\,$print("c")} is a shorthand for the continuation-passing style program%
\vspace{-2pt}
\begin{lstlisting}
	print("a", lambda x. print("b", lambda y. print("c", lambda z. ())))
\end{lstlisting}
\vspace{-8pt}
where \lstinline{print(s,k)} is a continuation-passing style version of \kw{print}, where we print to standard output and then explicitly call the continuation \kw{k}:
\lstinline{print(s,k) $\,\;$= print(s);k()}, though this behaviour may change based on an effect handler.

To simplify the lambda notation, consider the notation \lstinline{print(s;x.k)}, which explicitly says that the result of \lstinline{print(s)} is bound to \lstinline[basicstyle=\ttfamily]{x}, where the continuation \lstinline[basicstyle=\ttfamily]{k} depends on \lstinline{x}: \lstinline{print(x;s.k)} is equivalent to \lstinline{print(s, lambda x. k(x))}. The example above is then:
\vspace{-2pt}
\begin{lstlisting}
	print("a"; x. print("b"; y. print("c"; z. return$\footnotemark$ ())))
\end{lstlisting}
\vspace{-8pt}
\footnotetext{In this chapter, \lstinline[basicstyle=\ttfamily\footnotesize]{return x} can be thought of as an operation, whose default behaviour is to discard any subsequent continuation and finish the computation with \lstinline[basicstyle=\ttfamily\footnotesize]{x} (similarly to exceptions behaviour). A more in-depth discussion around the meaning of  \lstinline[basicstyle=\ttfamily\footnotesize]{return} is beyond the scope of this dissertation, but it is examined in detail in \cite[Section~1.2]{HillerstromPhDThesis}.}

As \lstinline{print} does not return a result, we may also simply write \lstinline{print(s. k)}:
\vspace{-2pt}
\begin{lstlisting}
	print("a". print("b". print("c". return ())))
\end{lstlisting}
\vspace{-8pt}

The \emph{handler} \lstinline{reverse} reverses the order in which \lstinline{print} operations are executed. Every \lstinline{print(s.k)} operation call inside of the context of \kw{reverse} is executed based on the body of the handler:  
the handler firstly resumes the continuation \lstinline{k}, and only then performs the operation itself. The computation \lstinline{with reverse handle abc} is the result of executing \lstinline{abc}, while handling operations with \lstinline{reverse}. This prints out `c', `b' and `a' in this order.

\vspace{-6pt}
\paragraph{Types of effect handlers.}

There exist different types of handlers, depending on the implementation, described in detail by \citet{Kammar2013} and \citet{Hillerstrom2016MSc}. For example, \textit{open handlers} forward unhandled operations to other handlers, but \textit{closed handlers} do not. Effects are handled only once with \textit{shallow handlers}, but they are propagated to any nested handlers and can be handled several times when using \textit{deep handlers}.
If the continuation \kw{k} must be invoked exactly once, then the handlers are \textit{linear},  while if it could be invoked more than once, they are \textit{multi-shot}. \textit{Exception handlers} do not invoke the continuation. 

Throughout this section, I assume open deep handlers. Edward2 uses open linear handlers, which are shallow, but allow for explicit effect forwarding, which mimics deep handlers.

\subsection{Composing effect handlers}
One very useful feature of open effect handlers is that they can be nested or composed to combine the way they interpret the computation. For example, consider a handler \kw{join}, which concatenates all strings appearing in a \kw{print} statement. One way to define such a handler is, as given by \citet{EffectsTutorial}:
\vspace{-4pt}
\begin{lstlisting}
	let join = 
		handler {
			return v $\quad\;\rightarrow$ (v, "")
			print(s. k) $\rightarrow$ let (v, acc) = k() in return (v, s + acc)$\footnotemark$ }
\end{lstlisting}
\vspace{-6pt}
The computation \lstinline{with join handle abc} evaluates to the tuple $((), \kw{"abc"})$. 
Using \kw{reverse} in the context of \kw{join} can be written as
\lstinline{with join handle with reverse handle abc},
which evaluates to  $((), \kw{"cba"})$. 
\footnotetext{\label{fn:string-concat}Here, $+$ is used to mean string concatenation.}

A more imperative-style way of writing handlers that pass around a ``state'' (in this case the accumulated string \kw{acc}) is with a mutable variable:
\vspace{-4pt}
\begin{lstlisting}
	set(acc, "");
	let join = handler { print(s. k) $\rightarrow$ set(acc, get(acc) + s); k() $\footnotemark[2]$ }
	with join handle with reverse handle abc;
	return acc
\end{lstlisting}
\vspace{-6pt}
Here, \kw{acc} is a mutable variable initialised to \kw{""} that we concatenate to at every \kw{print} statement. The final value of \kw{acc} holds the result: \kw{"cba"}.

This more imperative style is also closer to the Edward2 implementation of this chapter.
%
\subsection{Effect handling in probabilistic programming}
Recently, effect handlers have been adopted by some PPLs as a useful, modular way of implementing transformations of probabilistic programs for inference \cite{Scibior2015, Pyro, Edward2, ProbProg18, OliverMScDiss}.
The insight is to treat sampling statements as operations that can be handled by a separately defined handler. This enables a range of useful program transformations, including (though not limited to): conditioning, reparameterisation, tracing, density function derivation, variational family generation. When implemented in a differentiable programming framework, such as TensorFlow \cite{Tensorflow}, PyTorch \cite{PyTorch}, or JAX \cite{JAX}, the resulting code is also differentiable,%
\footnote{These frameworks do not typically provide native support for effects and handlers. The PPL effect-handling code is usually implemented in the differentiable language itself, meaning it is itself differentiable.}
making models easy to use in combination with gradient-based inference algorithms.

Consider the operation \kw{sample(dist; x. k)}, whose default behaviour is to sample at random from the distribution \kw{dist}, bind the result to the variable \kw{x} and invoke the continuation \kw{k}. 
Consider also the operation \kw{observed((value, dist); x. k)}, whose default behaviour simply binds \kw{x} to the value \kw{value} and invokes \kw{k}.
It is then straightforward to write a probabilistic model using this \kw{sample} operation. For example, the following code describes a Beta-Binomial model of $n$ trials:  
\begin{lstlisting}
	let beta_binomial(n) = 
		sample(beta(1., 1.); z);
		sample(binomial(z, n); x);
		return x
\end{lstlisting}
\vspace{-6pt}

To condition on some data, say $x = 7$, we can use an effect handler, which changes every \kw{sample} operation to an \kw{observed} operation:
\begin{lstlisting}
	let condition(rv, value) = 
			handler { sample(dist; rv. k) $\rightarrow$ observed((value, dist); rv); k(value) }
	let conditioned(n) = with condition(x,$\footnotemark$ 7) handle beta_binomial(n)
\end{lstlisting}
\vspace{-6pt}

\footnotetext{Here, $x$ is not a variable, but rather a \textit{name} of a variable. 
This is an unsatisfactory solution for the well-known problem of distinguishing between instances of the same effect \cite{daynight}.
How to refer to variables from outside of the model in a principled way, is a well-known problem also in probabilistic programming, where it leads to repeated syntax, such as Edward2's \lstinline[basicstyle=\ttfamily\footnotesize]{x = ed.Normal(0, 1, name="x")}.
}

This conditioning, however, is not enough for us to obtain a posterior of $z$. 
As we saw in \autoref{ch:background}, one way to perform inference is by repeatedly evaluating the joint probability density function of the model. Once again, we can use an effect handler to obtain the joint density function:
\begin{lstlisting}
	let joint(vals) = 
		set(density, 1);
		let h = handler { 
			sample(dist; rv. k) $\rightarrow$ 
					set(density, get(density) * pdf(dist,vals[rv])); 
					k(vals[rv])
			observed((value,dist); rv.k) $\rightarrow$ 
					set(density, get(density) * pdf(dist,value)); 
					k(value)
		}
		with h handle conditioned(n);
		return density
\end{lstlisting}
\vspace{-6pt}

Here \kw{joint} is a function on a dictionary that maps unobserved random variable names to a particular value in their domain. In the case above, we would evaluate \kw{joint} for \lstinline{vals = dict(z $\mapsto$ z_val)} for different values of \kw{z_val}.
Previously, we discussed handlers that explicitly handle only a single effect. Here, \lstinline{h} handles two effects, \lstinline{sample} and \lstinline{observed}, meaning instances of either of these effects will be handled in the context of \lstinline{h}.
The code also assumes that a function \kw{pdf} is available, such that for any distribution \kw{dist} and a value \kw{v} in the domain of \kw{dist}, \kw{pdf(dist, v)} evaluates the probability density function of \kw{dist} at \kw{v}.

Effect-handling based probabilistic programming languages have several advantages to density- and sampling-based PPLs. Conceptually, effect-handling based PPLs treat sampling abstractly and separate its use in defining the model from its actual implementation. Thus, models can be interpreted as either sampling-based, density-based, or in a completely different way, depending on usage. 
In addition, working with effects and handlers, and more specifically being able to compose them, allows for sophisticated program transformations, which do not need to be integrated into a separate compiler, but can be user-specified in the PPL itself. This highlights the potential of effect handlers as a basis for a programmable inference framework \cite{ProgrammableInference}, although explicit work has not yet been done in this direction.

The rest of this chapter shows several uses of effect handlers in the probabilistic programming language Edward2, presenting handlers that perform model reparameterisation and automatic variational guide synthesis.

\section{The paper} \label{sec:autoreparam-paper}   

The main contribution of this chapter is the paper \textit{Automatic Reparameterisation of Probabilistic Programs}. It considers the problem of \textit{reparameterisation} of probabilistic programs: changing the way the model is expressed in terms of parameters, in order to improve the quality of inference. The paper discusses the effects of parameterisation on the geometry of the posterior through examples, highlighting the practical challenges of finding a suitable reparameterisation. %
It presents two techniques for automatically reparameterising probabilistic programs using \textit{effect handlers}. The first is a simple interleaved sampling from two different model parameterisations. The second provides a novel continuous relaxation of the question of what parameterisation to use, and uses variational inference to find a suitable such parameterisation.

The paper was accepted for presentation at the \textit{Thirty-seventh International Conference on Machine Learning (ICML 2020)} and included in the \textit{Proceedings of Machine Learning Research, Volume 119}.
Out of $4990$ submissions in total, $1088$ papers were accepted.
An earlier version of the paper also appeared at the non-archival symposium \textit{Advances in Approximate Bayesian Inference (AABI 2018)}.

\paragraph{Author contributions.}
The paper is co-authored by me, Dave Moore and Matt Hoffman. 
My contributions included modifying Edward2 to enable composing effect handlers, performing the initial analytical analysis, developing the automatic reparameterisation framework, running preliminary experiments, and co-writing parts of the paper. 
Dave Moore mentored the project throughout, implemented improvements that allowed for using TensorFlow Probability's MCMC kernels more efficiently, ran the final experiments, did the additional analysis presented in Appendix D, and co-wrote parts of the paper.
Matt Hoffman conceived the idea of the project, mentored it throughout, co-wrote parts of the paper, and offered feedback, comments, and suggestions.

\newpage
\includepdf[pages=-,addtotoc={
1, subsection, 1, Introduction, p1,
2, subsection, 1, Related Work, p2,
2, subsection, 1, Understanding the Effect of Reparameterisation, p2,
3, subsection, 1, Reparameterising Probabilistic Programs, p3,
3, subsubsection, 1, Effect Handling-based Probabilistic Programming, p3,
4, subsubsection, 1, Model Reparameterisation Using Effect Handlers, p4,
5, subsection, 1, Automatic Model Reparameterisation, p5,
5, subsubsection, 1, Interleaved Hamiltonian Monte Carlo, p5,
5, subsubsection, 1, Variationally Inferred Parameterisation, p5,
7, subsection, 1, Experiments, p7,
7, subsubsection, 1, Models and Datasets, p7,
7, subsubsection, 1, Algorithms and Experimental Details, p7,
7, subsubsection, 1, Results, p7,
8, subsection, 1, Discussion, p8,
11, subsection, 1, Appendix A: Derivation of the Condition Number of the Posterior for a Simple Model, p11,
11, subsubsection, 1, Deriving $V_{CP}$ and $V_{NCP}$: Centred Parameterisation, p11,
11, subsubsection, 1, Deriving $V_{CP}$ and $V_{NCP}$: Non-centred Parameterisation, p11,
12, subsubsection, 1, The Best Diagonal Preconditioner, p12,
12, subsubsection, 1, The Condition Numbers $\kappa_{CP}$ and $\kappa_{NCP}$, p12,
12, subsection, 1, Appendix B: Interceptors, p12,
12, subsubsection, 1, Make log joint, p12,
12, subsubsection, 1, Non-centred Parameterisation Interceptor, p12,
13, subsubsection, 1, VIP Interceptor, p13,
13, subsubsection, 1, Mean-field Variational Model Interceptor, p13,
13, subsection, 1, Appendix C: Details of the Experiments, p13,
14, subsection, 1, Appendix D: Additional Analysis, p14
}]{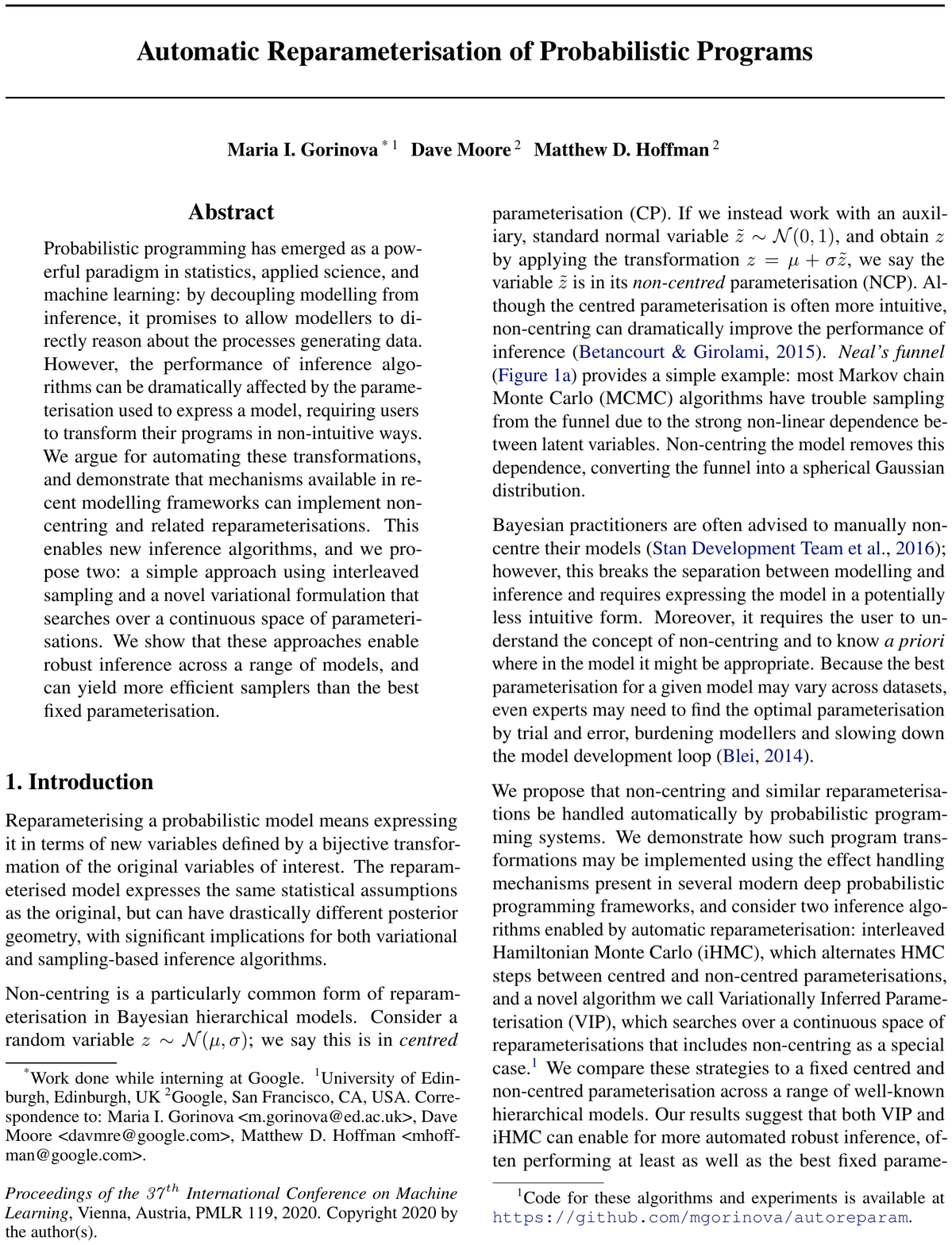}
\newpage

\section{Discussion} \label{sec:autoreparam-discuss}   

\subsection{Impact}

The work has been well received within other effect-handling based PPLs: both Pyro \cite{Pyro} and NumPyro \cite{NumPyro} have implemented the variationally inferred parameterisation (VIP) handler to allow for automatic reparameterisation.\footnote{\,{\notsotiny\url{https://docs.pyro.ai/en/latest/infer.reparam.html\#module-pyro.infer.reparam.loc_scale}}}
\textsuperscript{\hspace{-5pt},\hspace{-3pt}}
\footnote{\,{\notsotiny \url{http://num.pyro.ai/en/latest/reparam.html\#numpyro.infer.reparam.LocScaleReparam}}}

\subsection{Scope and extensions}
This chapter focused on one particular family of reparameterisations (the family of partial non-centred reparameterisations), applicable to one particular family of distributions (the location-scale family). However, it is possible to use effect handlers to perform model reparameterisation more generally. It is also possible to extend and adapt VIP to be applicable to other families of distributions and reparameterisations. 

\paragraph{Multivariate distributions.}
VIP can be used for a broad range of distributions: any univariate location-scale distribution (such as Gaussian, Logistic and Cauchy) are directly covered by the work of this paper. Adapting VIP to multivariate location-scale distributions is also possible. One way to do so for 
an $N$-dimensional 
variable $\boldsymbol{z} \sim \normal(\boldsymbol{\mu}, \Sigma)$ is to interpolate the eigenvalues of the covariance matrix $\Sigma$. If the eigendecomposition of $\Sigma$ is $\Sigma = Q\,\mathrm{diag}(\boldsymbol{\lambda})\,Q^{-1}$ (which also equals $Q\,\mathrm{diag}(\boldsymbol{\lambda})\,Q^T$, as $\Sigma$ is positive definite and symmetric, thus $Q$ is orthogonal), VIP can be implemented by interpolating $\tilde{\boldsymbol{\lambda}}$
between $\mathbf{1}$ and $\boldsymbol{\lambda}$ through the parameterisation parameters $\boldsymbol{\phi} \in [0, 1]^N$: we set $\tilde{\lambda}_i = \lambda_i^{\phi_i}$ for $i = 1, \dots, N$.
The reparameterisation is then:
$$\tilde{\boldsymbol{z}} \sim \normal(\tilde{\boldsymbol{\mu}},  \tilde{\Sigma}) 
\qquad\qquad
\mathbf{z} = \boldsymbol{\mu} + Q\,\mathrm{diag}(\sqrt{\boldsymbol{\lambda} \oslash \tilde{\boldsymbol{\lambda}}})\,Q^T\,(\tilde{\mathbf{z}} - \tilde{\boldsymbol{\mu}})$$

where 
$\tilde{\boldsymbol{\mu}} = \tilde{\boldsymbol{\lambda}} \odot \boldsymbol{\mu}$, 
$\tilde{\Sigma} = Q\,\mathrm{diag}(\tilde{\boldsymbol{\lambda}})\,Q^T$, 
$\odot$ denotes elementwise multiplication ($(\mathbf{a} \odot \mathbf{b})_i = a_i b_i$), and $\oslash$ denotes elementwise division ($(\mathbf{a} \oslash \mathbf{b})_i = a_i / b_i$).
As before, this results in interpolating between non-centred and centred parameterisation: when $\boldsymbol{\phi} = \mathbf{0}$, we have $\tilde{\boldsymbol{\mu}} = \mathbf{0}$ and $\tilde{\Sigma} = Q\,I\,Q^T = I$; when $\boldsymbol{\phi} = \mathbf{1}$, we have $\tilde{\boldsymbol{\mu}} = \boldsymbol{\mu}$ and $\tilde{\Sigma} = Q\,\mathrm{diag}(\boldsymbol{\lambda})\,Q^T = \Sigma$.

\paragraph{Other reparameterisations.}

Effect handlers can be used to implement reparameterisations more generally. In particular, if sampling a variable $z$ from a distribution $\mathrm{d1}$ is equivalent to sampling $\tilde{z}$ from a distribution $\mathrm{d2}$ and obtaining $z$ from $\tilde{z}$ via a transformation $f$, then we can define the following handler to reparameterise out models:
\begin{lstlisting}
	let reparam = handler {
		sample(rv; d1. k) $\rightarrow$ 
			sample(z_tilde; d2);
			let z = f(z_tilde);
			k(z) 
	}
\end{lstlisting}

For example, for a variable $z$ coming from any distribution $\mathrm{d}$ and provided we know $F_{\mathrm{d}}^{-1}$ --- the inverse cumulative distribution function (inverse CDF) of $\mathrm{d}$ --- we can reparameterise $z$ using an inverse CDF transform:
\begin{lstlisting}
	let icdf_reparam = handler {
		sample(rv; d. k) $\rightarrow$ 
			sample(z_tilde; uniform(0, 1));
			let z = $F_{\mathrm{d}}^{-1}$(z_tilde);
			k(z) 
	}
\end{lstlisting}

However, in order to adapt VIP to any new family of reparameterisations, we need to design a
continuous relaxation between the reparameterised/non-reparameterised model, the way VIP does so for non-centred/centred parameterisation. 
In other words, VIP is not immediately applicable to other reparameterisations, which is one of its biggest limitations.

%% file: chapters/conclusion.tex
\cleardoublepage
\bookmarksetup{startatroot}
\addtocontents{toc}{\bigskip}
\chapter{Challenges ahead}

Many challenges lie ahead of making Bayesian inference accessible. This thesis showed that probabilistic programming languages can utilise the underlying structure of a model to optimise inference. However, a lot of work remains before we can tailor a model-specific algorithm to a problem automatically.

Previous chapters already highlighted some of the future directions that remain unexplored. 
\autoref{ch:slicstan2} emphasised the need for a general framework for static analysis of probabilistic programs, which can slice a program according to a particular factorisation of interest, and perhaps facilitate programmable inference. 
\autoref{ch:autoreparam} demonstrated how to automate one specific model reparameterisation, but leaves the open questions of how to generalise to other reparameterisation families. 

Some other challenges of probabilistic programming were not mentioned here, but could be addressed by program analysis in the future. Perhaps most significantly, this dissertation did not consider problems where the number of model parameters is unbounded, nor inference algorithms that can be applied to such problems. Similarly, we assumed that the likelihood of the model is available in a closed form. Future work may look into factorising a model, so that challenging sub-parts of it, for example those containing an unbounded number of variables, or deterministic observations, are treated separately to automatically synthesise an efficient model-specific strategy.